\documentclass[10pt,twocolumn,letterpaper]{article}

\usepackage{cvpr}
\usepackage{times}
\usepackage{epsfig}
\usepackage{graphicx}
\usepackage{amsmath}
\usepackage{amssymb}

\usepackage{comment}
\usepackage{subcaption}
\usepackage{algorithm}
\usepackage{algorithmicx}
\usepackage{algpseudocode}
\usepackage{commath}

\usepackage{soul,color}
\definecolor{Red}{cmyk}{0,1,1,0}
\definecolor{Green}{cmyk}{1,0,1,0}
\definecolor{Cyan}{cmyk}{1,0,0,0}
\definecolor{Purple}{cmyk}{0.45,0.86,0,0}
\definecolor{Rosolic}{cmyk}{0.00,1.00,0.50,0}
\definecolor{Blue}{cmyk}{1.00,1.00,0.00,0}
\definecolor{Orange}{cmyk}{0,0.52,0.80,0}
\definecolor{Black}{cmyk}{1,0,0,1}

\newcommand{\hb}[1]{{\color{Black}            {#1}}}

\usepackage[pagebackref=true,breaklinks=true,letterpaper=true,colorlinks,bookmarks=false]{hyperref}

\cvprfinalcopy 
\def\cvprPaperID{8148} 

\ifcvprfinal\fi
\begin{document}

\title{Fast 3D Indoor Scene Synthesis \\ with Discrete and Exact Layout Pattern Extraction}

\author{Song-Hai Zhang\\
Tsinghua University\\
{\tt\small shz@tsinghua.edu.cn}
\and
Shao-Kui Zhang\\
Tsinghua University\\
{\tt\small zhangsk18@mails.tsinghua.edu.cn}
\and
Wei-Yu Xie\\
Beijing Institute of Technology\\
{\tt\small ervinxie@qq.com}
\and
Luo-Cheng Yang\\
Tsinghua University\\
{\tt\small luocy16@mails.tsinghua.edu.cn}
\and
Hong-Bo Fu\\
City University of Hong Kong\\
{\tt\small fuplus@gmail.com}
}

\maketitle

\begin{abstract}
We present a fast framework for indoor scene synthesis, given a room geometry and a list of objects with learnt priors. Unlike existing data-driven solutions, which often extract priors by co-occurrence analysis and statistical model fitting, our method measures 
the strengths of spatial relations by tests for complete spatial randomness (CSR), and extracts complex priors based on samples with the ability to accurately represent 
discrete layout patterns. With the extracted priors, our method achieves both acceleration and plausibility by partitioning input objects into disjoint groups, followed by layout optimization
based on the Hausdorff metric.
Extensive experiments show that our framework is capable of measuring more reasonable relations among objects and simultaneously generating varied arrangements in seconds. 
\end{abstract}

\section{Introduction} \label{sec:introduction}
3D indoor scene synthesis is thriving in recent years. As demonstrated in  \cite{asurveyof3didss,germer2009procedural,lyons2008ten}, automatically synthesizing plausible rooms benefits various applications. 
With the emergence of various datasets for 3D indoor scenes \cite{song2016ssc,InteriorNet18,Structured3D}, techniques shift toward data-driven approaches, i.e., modeling priors expressing strategies of layouts of furniture objects. However, inherent difficulties of 3D indoor scene synthesis still exist in various aspects. 

\begin{figure}[!t]
	\centering
	\begin{subfigure}[b]{0.32\linewidth}
		\includegraphics[width=\linewidth]{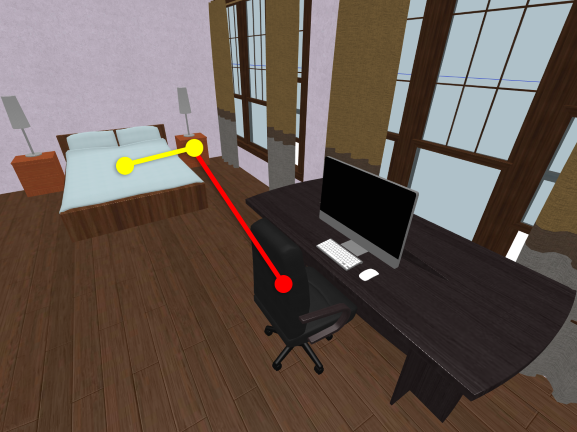}
		\caption{A bedroom. }
		\label{fig:cooccur}
	\end{subfigure}
	\begin{subfigure}[b]{0.32\linewidth}
		\includegraphics[width=\linewidth]{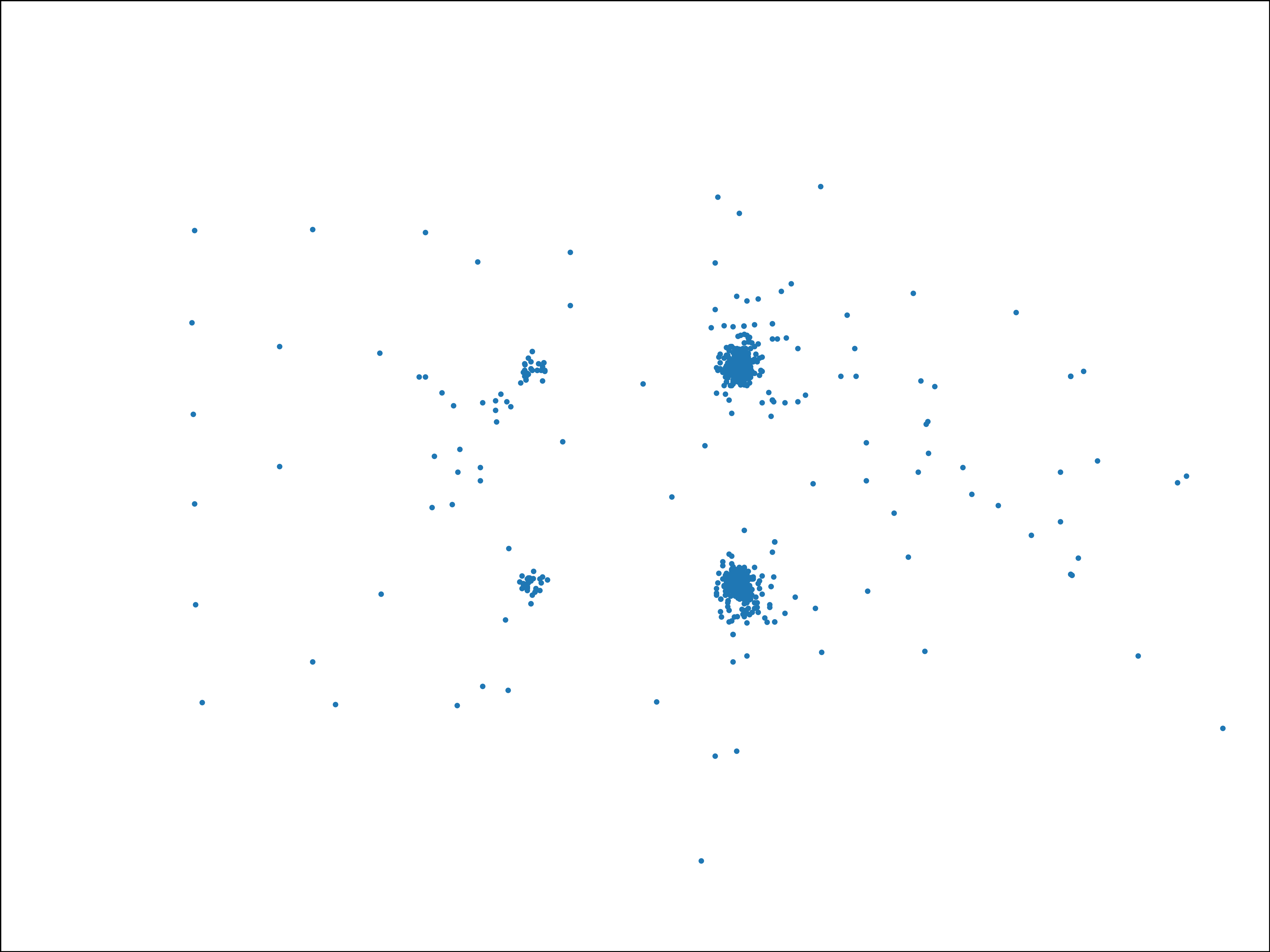}
		\caption{A stand \& a bed.}
		\label{fig:cooccur2}
	\end{subfigure}
	\begin{subfigure}[b]{0.32\linewidth}
		\includegraphics[width=\linewidth]{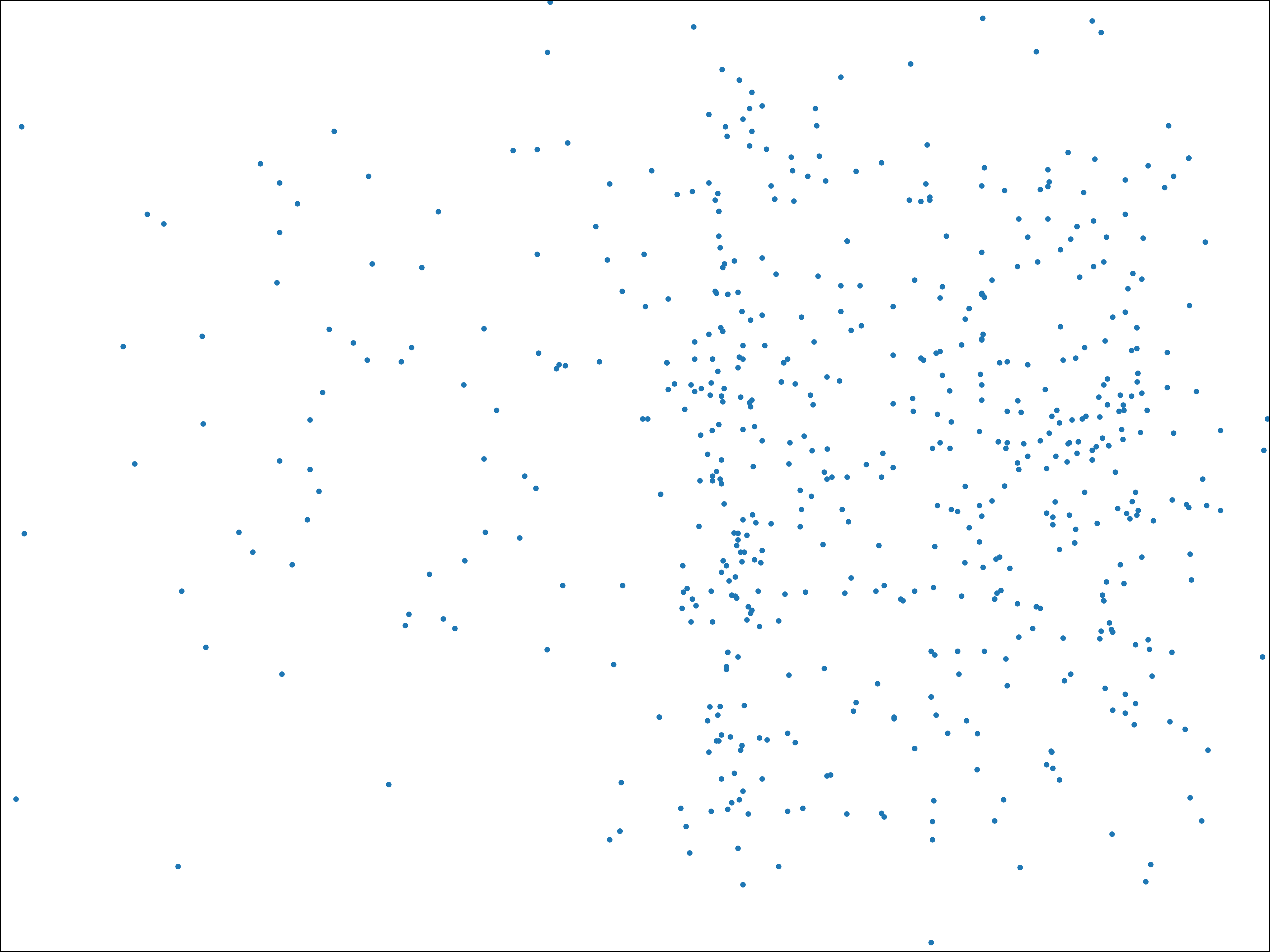}
		\caption{A stand \& a chair.}
		\label{fig:cooccur3}
	\end{subfigure}

	\caption{\hb{Illustrating} the problems of co-occurrence. With similar co-occurrence, two relative positions of two pairs of objects are shown in \ref{fig:cooccur}. In \ref{fig:cooccur2}, the double bed and the stand are obviously spatial related, while the stand and the chair are distributed randomly. }
	\label{fig:co-occur-problem}
\end{figure}

First, it is inevitable to deal with furniture layouts 
parameterized continuously or discretely, which distribute in complex high-dimensional spaces \cite{li2018df}. A few works (e.g., \cite{fu2017adaptive,qi2018human,fisher2012example,ma2018language}) attempt to simplify layouts 
into independent cliques or subsets e.g., \cite{fisher2012example,qi2018human}. However, their underlying metric depends on ``co-occurrence", which is merely counting co-existing frequencies instead of incorporating spatial knowledge. For an example in Figure \ref{fig:co-occur-problem}, a high frequency of co-occurrence does not necessarily signify a strong spatial relationship. In other words, scene synthesis purely based on co-occurrence could generate weird outcomes. 

\begin{figure}[t]
    \centering
    \includegraphics[width=\linewidth]{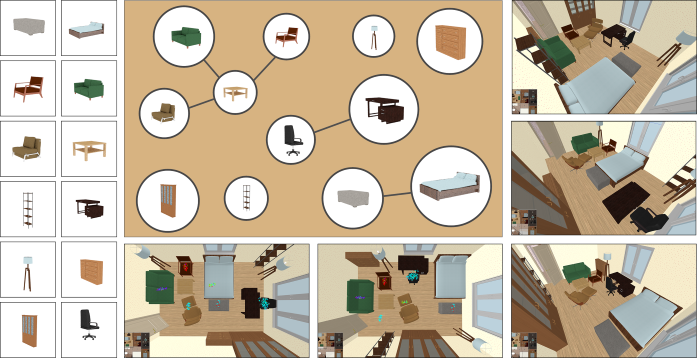}
    \caption{Given a list of objects (Left), we decompose them into disjoint groups (Top-Middle) with coherence for each individual group and freedom among groups. 
    By incorporating discrete templates as priors to guide syntheses, our method generates 
    various plausible layouts in seconds. 
    }
    \label{fig:teaser}
\end{figure}

\begin{figure}[!ht]
	\centering
	\begin{subfigure}[b]{0.32\linewidth}
		\includegraphics[width=\linewidth, scale=0.5]{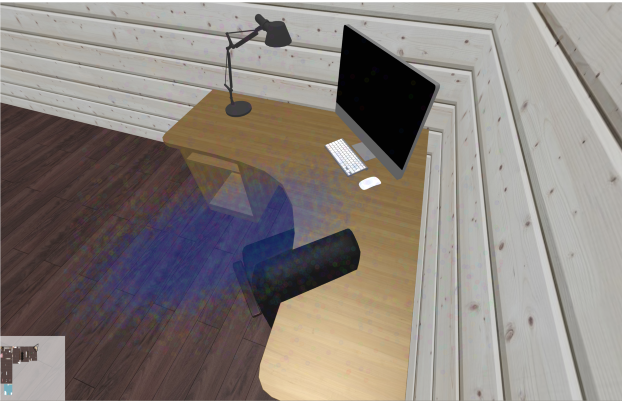}
		\caption{GMM.}
		\label{fig:inherient3}
	\end{subfigure}
	\hfill
	\begin{subfigure}[b]{0.32\linewidth}
		\includegraphics[width=\linewidth, scale=0.5]{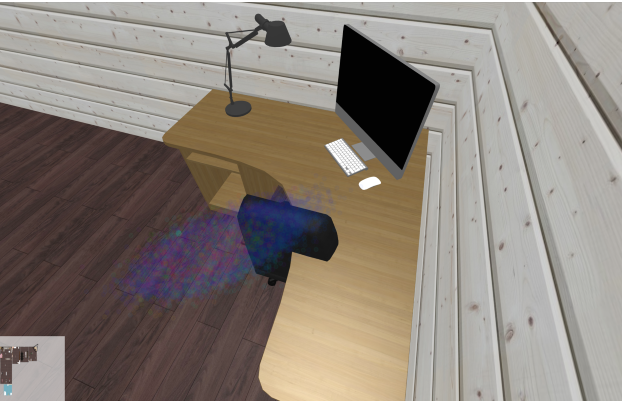}
		\caption{GMM fits ours. }
		\label{fig:inherient2}
	\end{subfigure}
	\hfill
	\begin{subfigure}[b]{0.32\linewidth}
	    \includegraphics[width=\linewidth, scale=0.5]{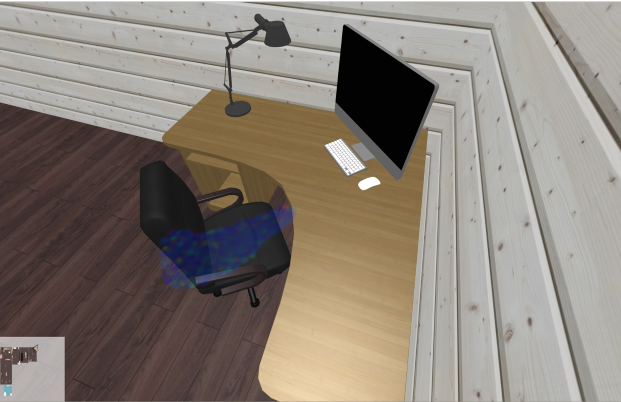}
	    \caption{Ours. }
	    \label{fig:inherient1}
	\end{subfigure}
	\caption{Inherent problems of fitting models. Fitting raw data (Left) suffers with noises. GMM fitting our denoised data (Middle) gives a blurred pattern. Ours (Right) successfully places the chair at the right place in a correct orientation. }
	\label{fig:inherient}
\end{figure}

Second, due to innumerable strategies of arrangement, it is hard to exhaustively list all possible spatial relations among objects \cite{chang2015text,chang2014learning,merrell2011interactive,yeh2012synthesizing,li2019grains} or to mathematically formulate unified and accurate models for them \cite{fisher2012example,xu2013sketch2scene,wang2019planit,wang2018deep}. For example, Chang et al. \cite{chang2015text} dictate a specific set of possible relations such as ``support", ``right", ``front", etc, which fundamentally limit the variety of possibly synthesized scenes. To model relations with multiple patterns, a common approach is to fit observed layouts with models. However, ``fitting models" could potentially introduce noises and be influenced by noises, especially when the underlying patterns do not satisfy with the assumptions of models, e.g., a commonly used Gaussian mixture model (GMM). Figure \ref{fig:inherient} shows a failure case of sampling a relative position and an orientation from a GMM of a chair \textit{w.r.t} a table. We argue that when the observed data is of sufficient size, the correct case inside observations or samples for fitting already offers exact layout strategies with varieties.




To address the above difficulties, in this paper, we propose a method to measure the strength of spatial relations between objects by utilizing tests for complete spatial randomness (CSR) \cite{diggle1979parameter}. A test for CSR (Section \ref{sec:csr}) describes how likely a set of events are generated \textit{w.r.t} a homogeneous Poisson process. Intuitively, it measures how obvious certain patterns exist in a set of points. Therefore, objects with high measurements tend to be grouped and arranged together. Objects that fail to pass tests for CSR are ignored, even if they have high co-occurrence. 

Furthermore, we present an approach for extracting discrete representation of various shapes of layout strategies, incorporating density peak clustering \cite{rodriguez2014clustering}. Finally, we present a framework for automatically synthesizing various arrangements of given objects \textit{w.r.t} an input room geometry, by partitioning input objects into disjoint groups according to the extracted priors, followed by optimization based on the Hausdorff metric to cope with discrete priors. The entire process can be done in seconds.

In sum our work makes the following contributions: 
\begin{enumerate}
	\item We first incorporate tests for complete spatial randomness to measure the strength of spatial relations between objects, which is a more powerful measurement than ``co-occurrence", allowing us to decompose a given list of objects into several disjoint groups. Therefore, unnecessary calculations are reduced whilst the plausibility is increased for indoor scene modeling.  
	\item We propose to use discrete and exact distributions to represent layout patterns of arbitrary shapes of objects for indoor scene configurations.
	\item We introduce a fast indoor scene synthesis framework, which is able to generate diverse arrangements in seconds.
\end{enumerate}

\section{Related Works}
\textbf{3D Indoor Scene Synthesis} aims at generating appropriate layouts of furniture objects for rooms. Various solutions considering different input settings and tasks have been proposed. For example, \cite{avetisyan2018scan2cad,chen2014automatic,fisher2015activity,shao2012interactive} generate room layouts based on RGB-D images or 3D scans. \hb{Human language \cite{chang2014learning,chang2015text,ma2018language}} and hand-drawn sketches \cite{xu2013sketch2scene} have also been explored as additional inputs to guide scene synthesis. \cite{wang2018deep,ritchie2019fast,ma2016action} iteratively infer the next objects to rooms. A full review of existing works on indoor scene synthesis is beyond the scope of this paper. Please refer to an insightful survey in \cite{asurveyof3didss}. 

As discussed in Section \ref{sec:introduction}, representations of layout strategies play an important role in 3D indoor scene synthesis. To encode prior knowledge, \cite{yeh2012synthesizing,merrell2011interactive,weiss2018fast} \hb{attempt to quantify interior} design rules. The emerging availability of 3D indoor scene datasets \hb{enables various} data-driven approaches. \hb{For example,} \cite{wang2019planit,chang2014learning} model \hb{spatial} relations between objects using semantics such as ``left", ``right", ``front", etc. Gaussian mixture models (GMMs) are adopted by \cite{fisher2012example,xu2013sketch2scene,henderson2017automatic} to fit observed distributions of objects. Graph structures are constructed by \cite{qi2018human,fu2017adaptive}. As surveyed in \cite{asurveyof3didss}, \cite{yu2011make,liang2018knowledge} model contexts for objects, e.g., average orientations and distances between objects, orientations \textit{w.r.t} the nearest walls, etc. However, despite the variety of representations, the underlying metrics are still confined to co-occurrence,  model fitting or even intuitive semantics, e.g., probabilities of edges are calculated by co-existing frequencies \cite{fisher2012example,qi2018human}. 

Our task partially resembles \cite{yu2011make} and \cite{weiss2018fast}, \hb{but takes an automatic approach to extract constraints from existing layouts.} \hb{We are also inspired by the works of} \cite{fisher2012example} and \cite{xie2013reshuffle}. \hb{However, the former requires exemplar scenes as input, while the latter focuses on re-arrangement of existing scenes. In contrast, we aim to learn general patterns for pairs of objects from existing layout examples for scene synthesis.}

\textbf{Tests for Complete Spatial Randomness} (CSR) is a classical topic \cite{diggle1983statistical}. Given a series of points distributed on a plane, a test for CSR is typically \hb{used to answer} how likely the points are placed randomly. Formally, it \hb{describes} how likely a set of events are generated \textit{w.r.t} a homogeneous Poisson process (planar Poisson process). Previously, most applications of CSR are confined to ecology \cite{gignoux1999comparing}, e.g., to investigate whether or not a set of observed plants are located with patterns. Rosin \cite{rosin1998thresholding} \hb{is probably the first to bring the}  concept of CSR into computer vision to handle the problem of how to detect white noises inside images. Typical methods of tests for CSR include using Diggle's function \cite{diggle1979parameter,ho2006testing}, distance-based methods \cite{diggle1983statistical}, etc. In this paper, we \hb{follow \cite{assuncao1994testing} to test} CSR by means of angles (Section \ref{sec:csr}). 

\begin{figure}[!ht]
	\centering
	\begin{subfigure}[b]{0.32\linewidth}
		\includegraphics[width=\linewidth]{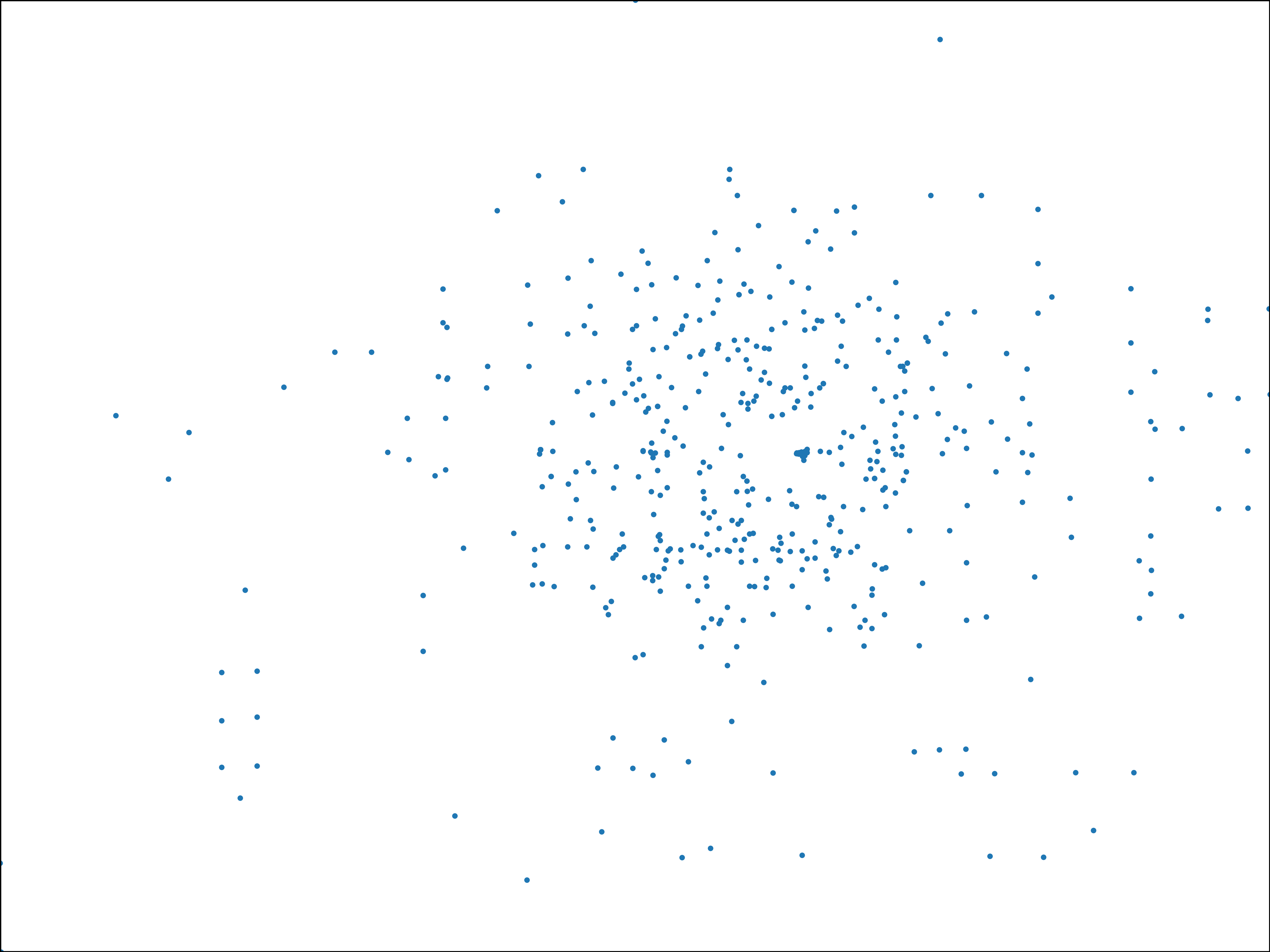}
		\caption{$d^{wa,ct}=1.12$.}
		\label{fig:resultfewcsr_1}
	\end{subfigure}
	\begin{subfigure}[b]{0.32\linewidth}
		\includegraphics[width=\linewidth]{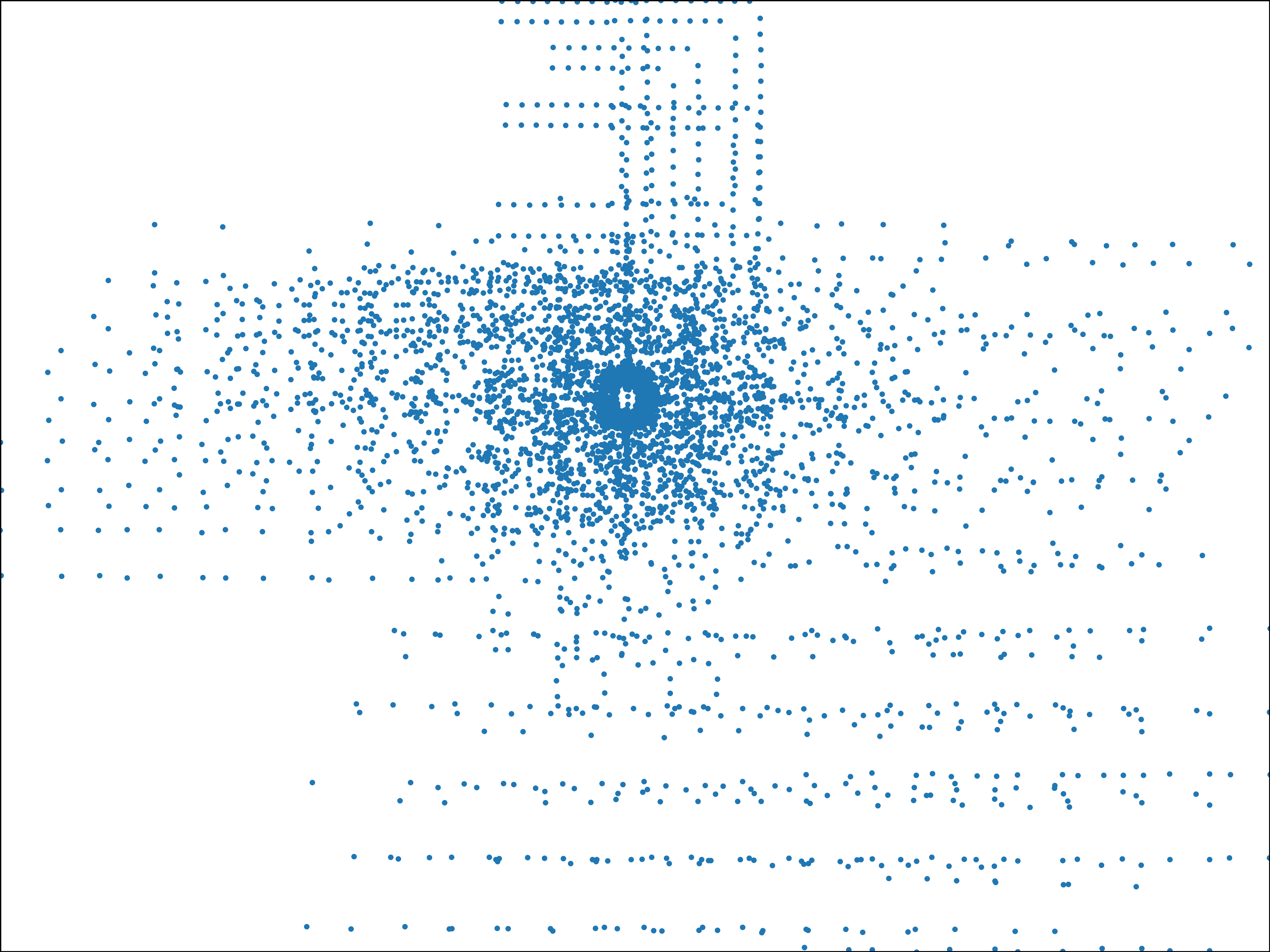}
		\caption{$d^{dt,ch}=2.03$.}
		\label{fig:resultfewcsr_2}
	\end{subfigure}
	\begin{subfigure}[b]{0.32\linewidth}
		\includegraphics[width=\linewidth]{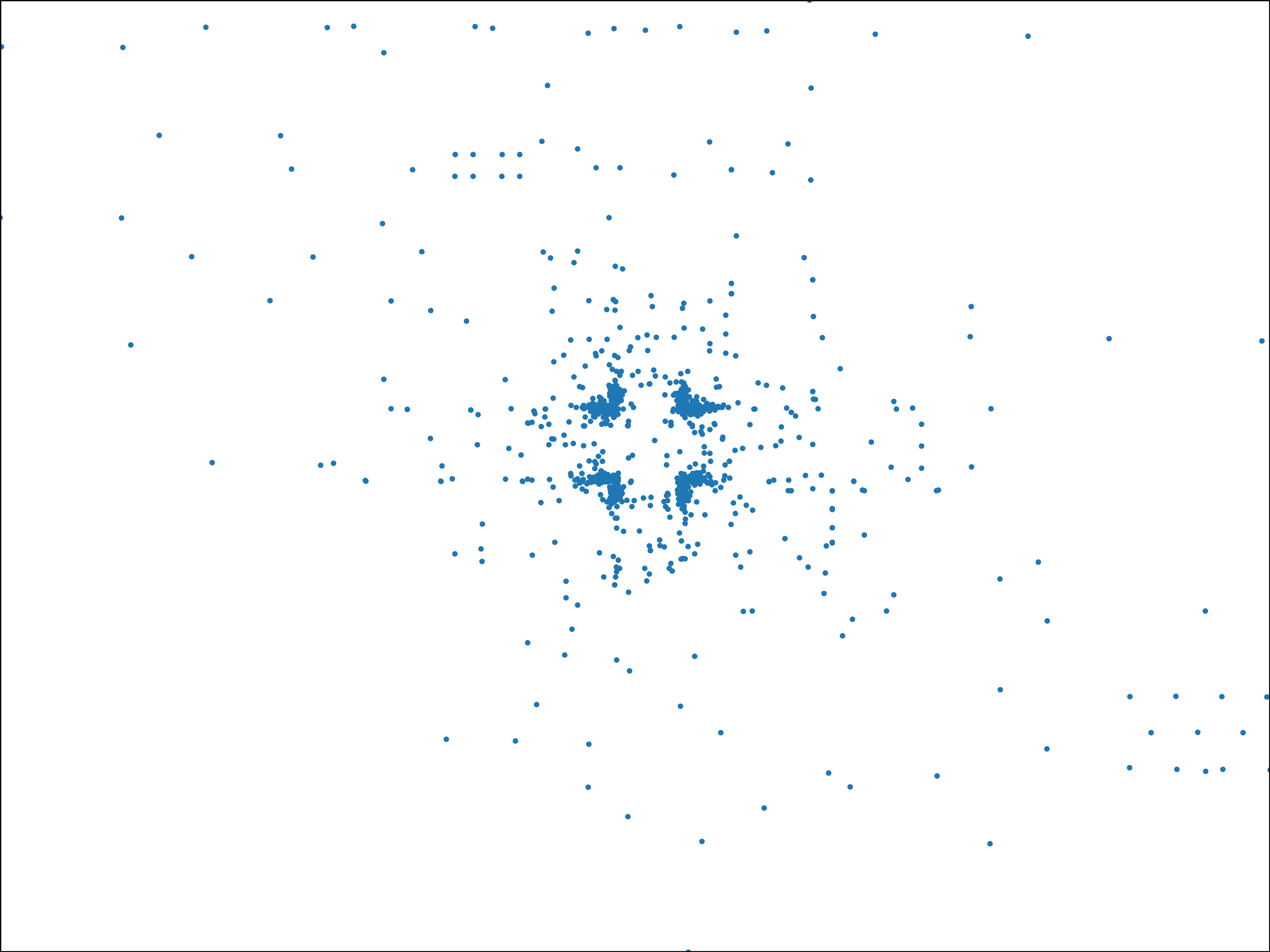}
		\caption{$d^{be,ni}=2.47$.}
		\label{fig:resultfewcsr_3}
	\end{subfigure}	
	\caption{Several results of tests for CSR. \ref{fig:resultfewcsr_1} plots relative positions between a wardrobe cabinet and a coffee table, \ref{fig:resultfewcsr_2} plots relative positions between a dinning table and a chair, and  \ref{fig:resultfewcsr_3} plots relative positions between a bed and a nightstand. }
	\label{fig:resultfewcsr}
\end{figure}

\section{Overview} \label{sec:overview}	
Our pipeline is split into an offline stage and an online stage. In the offline stage, we first learn the spatial strength graph $G_{ss}$ indicating how objects are spatially related with each other (section \ref{sec:csr}). This is more powerful than counting co-occurrence. We also extract versatile patterns of layout strategies as discrete ``templates" and reduce noises within datasets such as SUNCG \cite{song2016ssc} (Section \ref{sec:pe}). 
Given learned priors, an empty room, and a set of user-specified objects, 
during the online stage, our method first groups spatially coherent objects into groups (e.g., a bed and two night stand, as illustrated in Figure \ref{fig:srg_abstract}). 
Next, we do an instant arrangement for each group by heuristically using learned templates. Finally, we adjust the overall layout by optimizing a consistent loss function (Section \ref{sec:online}). 

In formal, the offline input is a multigraph $G_{in}=(V_{in},E_{in})$, which is a direct mathematical representation of original datasets, i.e., each vertex corresponds to an objects and each edge corresponds to a displacement between two objects. A vertex $v_{in}^{i} \in V_{in}$ contains a set of attributes $\{(d_{wall}^{i,\omega}, \theta_{wall}^{i,\omega}, t_{wall}^{i, \omega}) | \omega = 1,2,3 \dots, \Omega\}$, i.e., the row values of distances, orientations and translations \textit{w.r.t} its nearest walls.

Centering an object $o_i$, the $k$-th edge $e_{in}^{i,j,k} \in E_{in}$ from $v_{in}^{i}$ to $v_{in}^{j}$ is valued by a quadruple $(p_x^{i,j,k},p_y^{i,j,k},p_z^{i,j,k},p_\theta^{i,j,k})$ representing the $k$-th relative translation and orientation of $o_j$ \textit{w.r.t} $o_i$. 
And we leverage $E_{in}^{i,j}$ to indicate the set of edges formed from $v_{in}^{i}$ to $v_{in}^{j}$, where $v_{in}^{i}$ is the corresponding vertex in $V_{in}$ of object $o_i$. 

We construct $G_{in}$ with 2266 vertices, over 2 million edges from more than 520,000 rooms in SUNCG dataset, and measure their strength of spatial relations in Section \ref{sec:csr} and extract layout priors in Section \ref{sec:pe}. 


\begin{figure}[!ht]
	\centering
	\begin{subfigure}[b]{0.58\linewidth}
		\includegraphics[width=\linewidth]{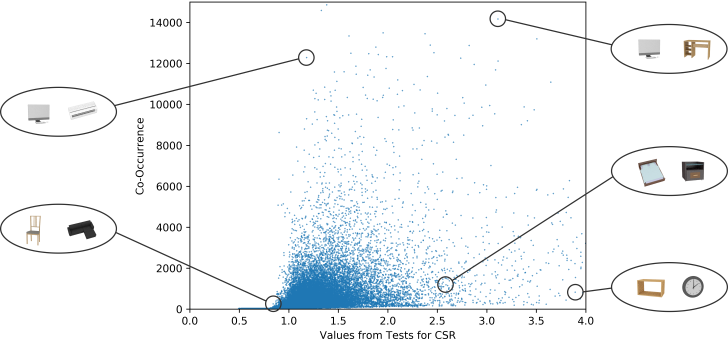}
		\caption{ }
		\label{fig:csrall}
	\end{subfigure}
	\hfill
	\begin{subfigure}[b]{0.38\linewidth}
		\includegraphics[width=\linewidth]{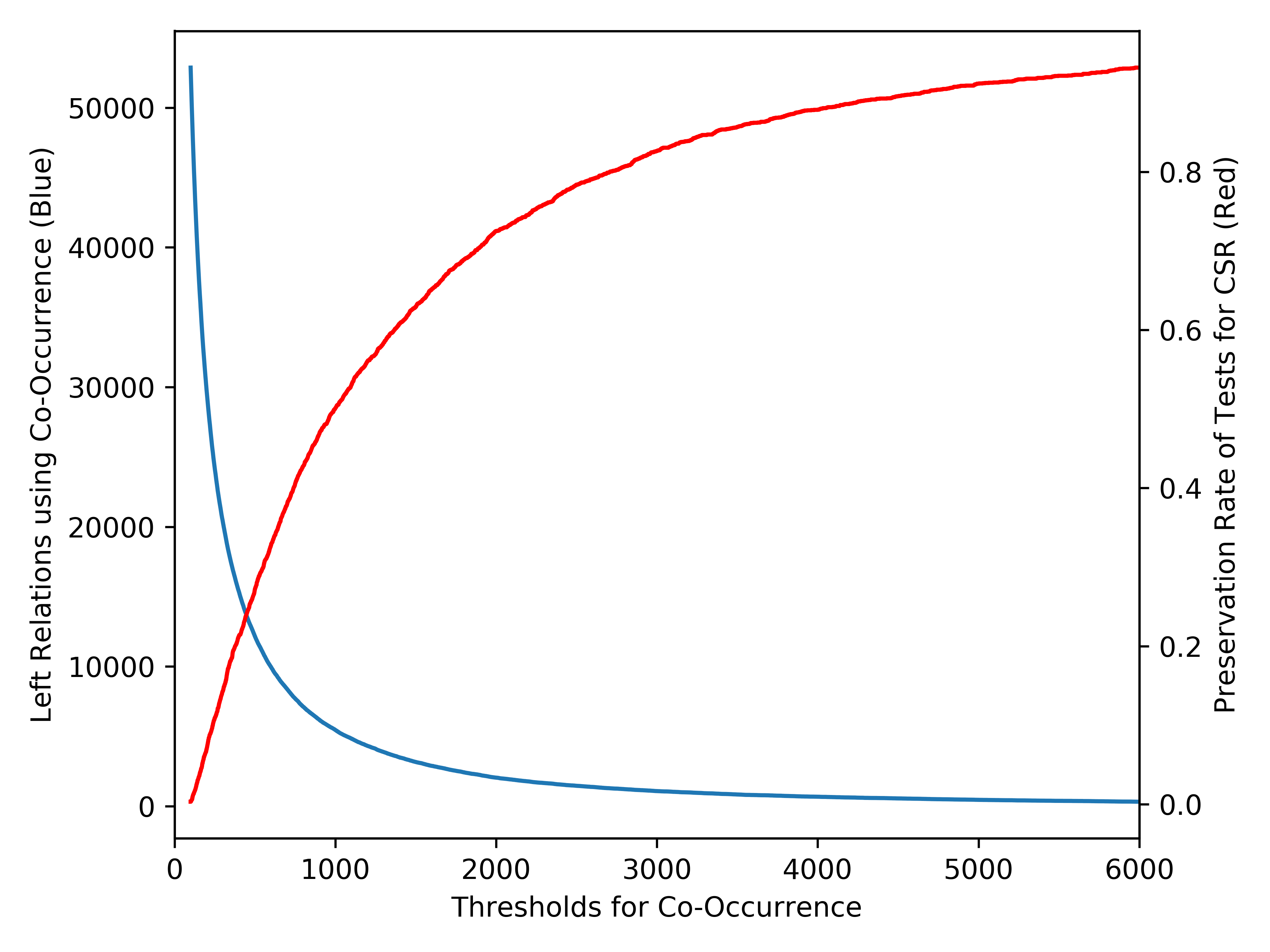}
		\caption{  }
		\label{fig:thresholds}
	\end{subfigure}
	\caption{\ref{fig:csrall} plots the CSR value and co-occurrence of \hb{every pair of objects}. Two objects might co-occur in many rooms, while the strength of their spatial relation could be low, vice versa. For example, the bed and the nightstand do have patterns of each other despite their low co-occurrence. \ref{fig:thresholds} plots how tests for CSR are able to retain relations which are mistakenly removed by co-occurrence with increasing thresholds set for co-occurrence. }
	\label{fig:csranalysis}
\end{figure}

\section{Spatial Strength Graph} \label{sec:csr}
Before actually extracting a template from datasets for each pair of objects, a question naturally arises: do we require templates for all pairs? As shown in Figure \ref{fig:co-occur-problem}, two objects could have very messy layout strategies, with transformations between them rather independent of each other, even though they might have high co-occurrence. This motivates us to learn a spatial strength graph (SSG) so that a multitude of pairs of objects that have low relations of spatial strength are ignored when arranging rooms. This will help us synthesize more plausible scenes but also accelerate the synthesis process. 

Formally, an SSG is a weighted graph defined as $G_{ss}=(V_{ss},E_{ss})$, where $G_{ss}$ denotes an entire graph, with $V_{ss}=V_{in}$ representing all objects in the dataset, and $E_{ss}$ is the edges with weight to encode spatial strength between objects. 
We measure the weights of $E_{ss}$ by equation \ref{equ:dvalue} that is ``d-value" \cite{assuncao1994testing} within the domain of tests for complete spatial randomness (CSR) \cite{diggle1979parameter}:
\begin{equation} \label{equ:dvalue}
d=\sqrt{m}\sup|F_{c}(\theta)-F_{e}(\theta)|. 
\end{equation}
$F_{c}$ and $F_{e}$ are respectively cumulative distribution function (CDF) and empirical distribution function (EDF) \textit{w.r.t} angle $\theta$, which is subject to uniform distribution \cite{assuncao1994testing}. $m$ is the number of points formulating the $F_e$.
For each pair of objects $o_i$ and $o_j$, the weights $E_{ss}^{i,j}$ is set to $d^{i,j}$ subject to random samples from $E_{in}^{i,j}$ in a ratio of 10\%, as suggested in \cite{assuncao1994testing} and \cite{diggle1976statistical}.
As shown in figure \ref{fig:resultfewcsr_1}, a wardrobe and a coffee table are spatilly independent, so their d-value is low. Although considerable noises exist in figure \ref{fig:resultfewcsr_2}, d-value of a dinning table and a chair is still reasonablly high. Finally, figure \ref{fig:resultfewcsr_3} shows clear patterns between a bed and a nightstand. 

Figure \ref{fig:csranalysis} suggests the differences between tests for CSR and co-occurrences. In figure \ref{fig:csrall}, we plot two measurements for all pairs of objects, where pairs including an air-conditioner typically co-occur frequently but air-conditioners are placed independently to most of other objects. Figure \ref{fig:thresholds} illustrates how tests for CSR are able to retain relations mistakenly removed by co-occurrences,

\section{Prior Extraction} \label{sec:pe}

\begin{figure}[!ht]
	\centering
	\begin{subfigure}[b]{0.32\linewidth}
		\includegraphics[width=\linewidth]{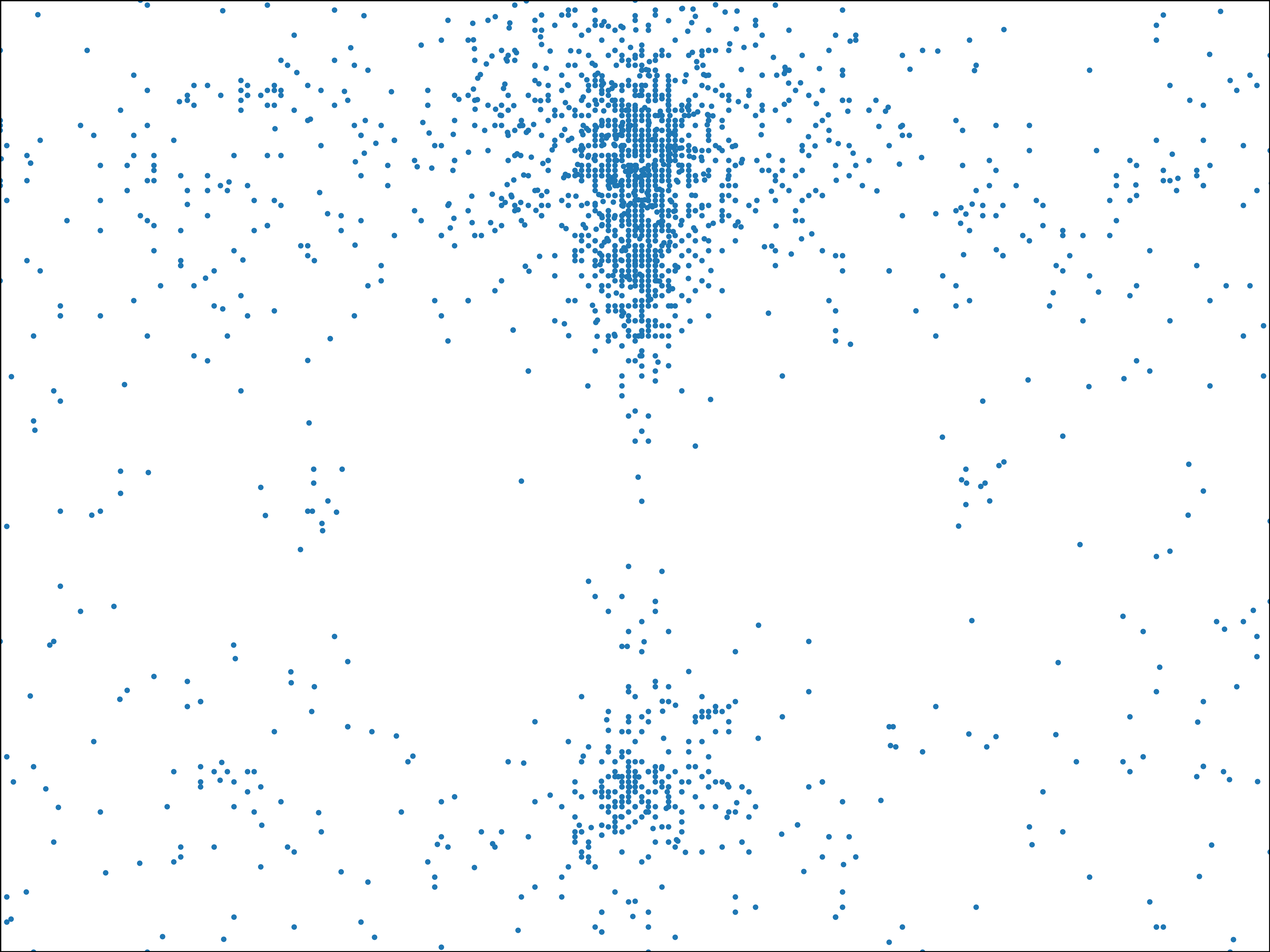}
		\caption{Input. }
		\label{fig:pe_origin}
	\end{subfigure}
	\begin{subfigure}[b]{0.32\linewidth}
		\includegraphics[width=\linewidth]{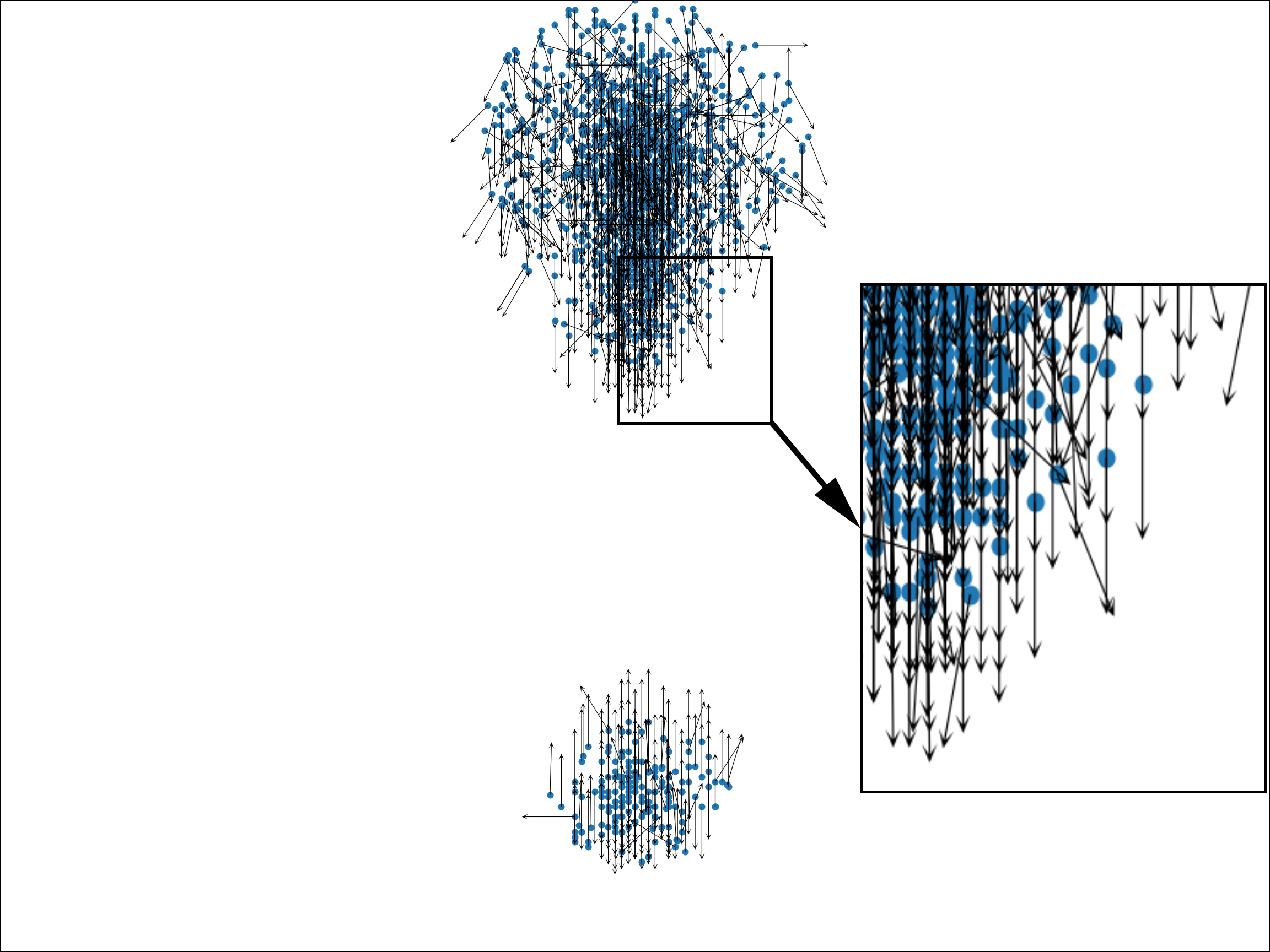}
		\caption{De-noised. }
		\label{fig:pe_denoised}
	\end{subfigure}
	\begin{subfigure}[b]{0.32\linewidth}
		\includegraphics[width=\linewidth]{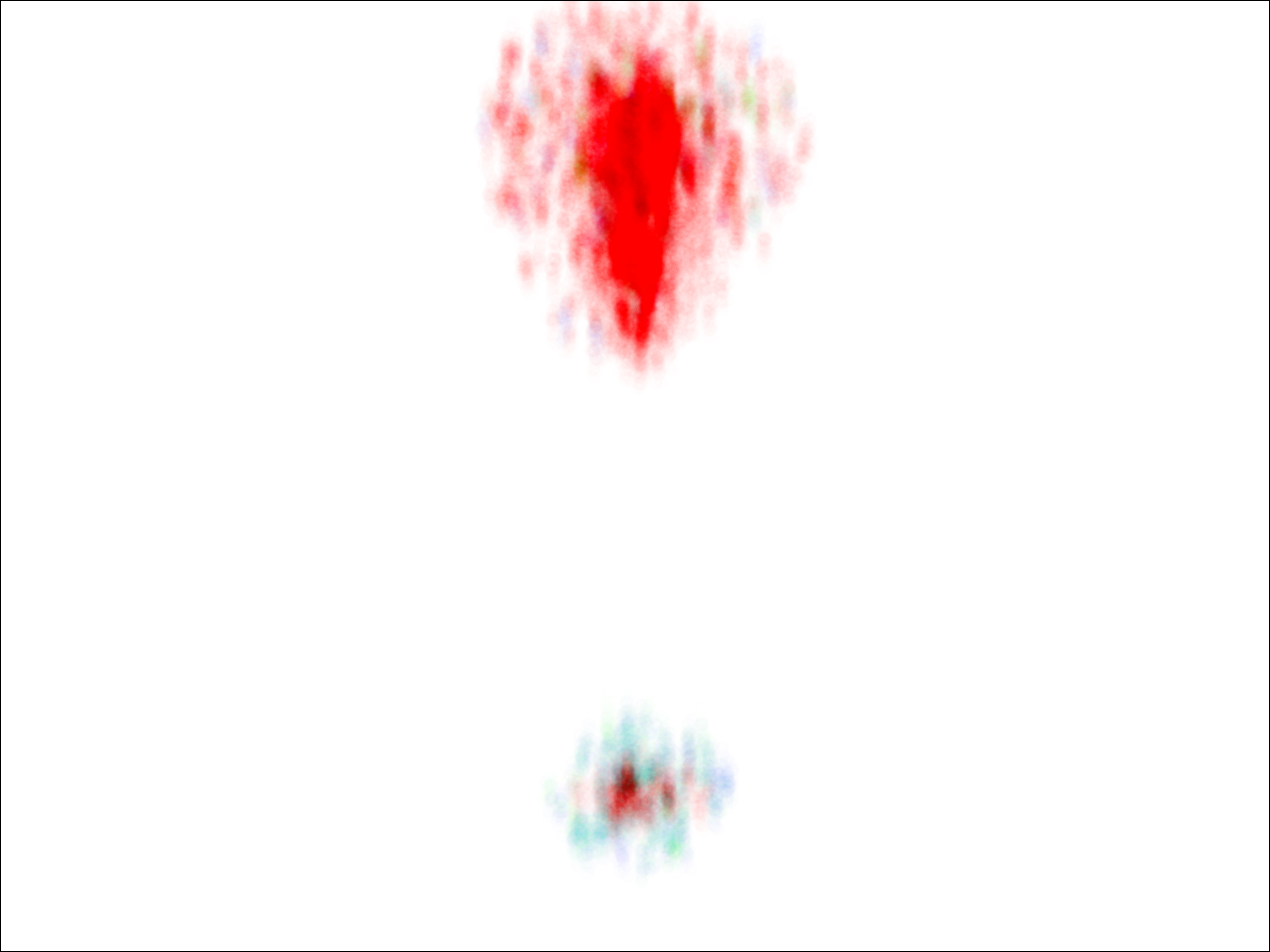}
		\caption{Fitted. }
		\label{fig:pe_fitted}
	\end{subfigure}
	\caption{The overall process of prior extraction. \ref{fig:pe_origin} \hb{is the input with} considerable noises. \ref{fig:pe_denoised} de-noises the input, which is readily to use in our framework, while \ref{fig:pe_fitted} depicts the further generalization of our templates into fitted models which are applicable for other frameworks such as MCMC }
	\label{fig:pe_odf}
\end{figure}

\begin{figure*}[!ht]
	\centering
	\begin{subfigure}[b]{0.16\linewidth}
		\includegraphics[width=\linewidth]{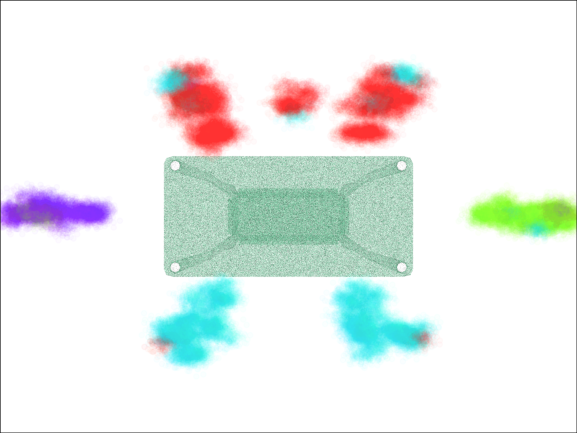}
		\caption{CoffeeTable-Chair}
	\end{subfigure}
	\begin{subfigure}[b]{0.16\linewidth}
		\includegraphics[width=\linewidth]{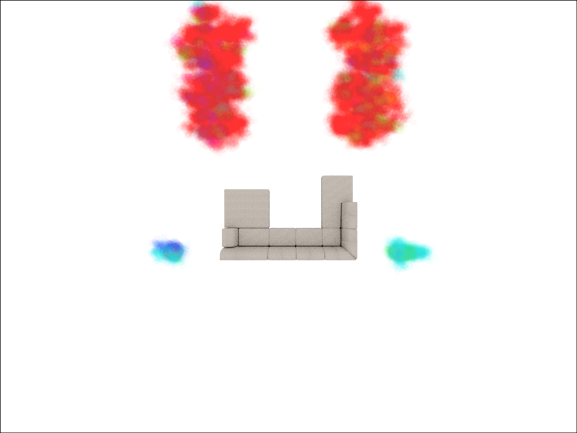}
		\caption{Sofa-Loudspeaker}
	\end{subfigure}
	\begin{subfigure}[b]{0.16\linewidth}
		\includegraphics[width=\linewidth]{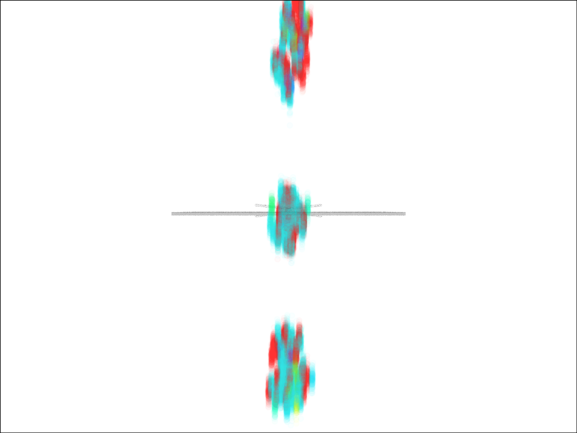}
		\caption{TV-CoffeeTable}
	\end{subfigure}
	\begin{subfigure}[b]{0.16\linewidth}
		\includegraphics[width=\linewidth]{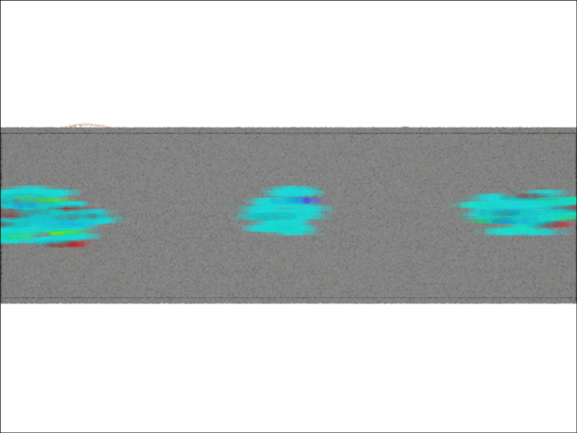}
		\caption{{\scriptsize TVStand-PlayStation}}
	\end{subfigure}
	\begin{subfigure}[b]{0.16\linewidth}
		\includegraphics[width=\linewidth]{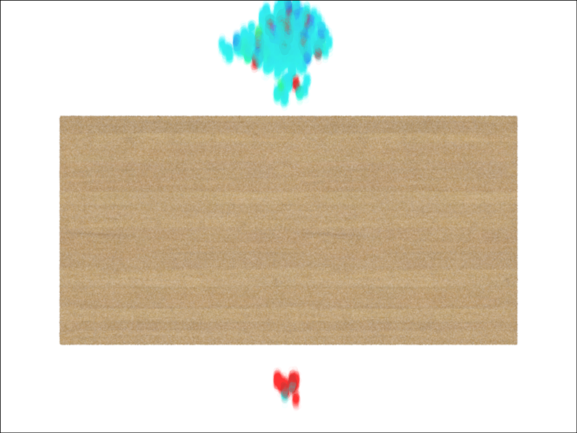}
		\caption{Desk-OfficeChair}
	\end{subfigure}
	\begin{subfigure}[b]{0.16\linewidth}
		\includegraphics[width=\linewidth]{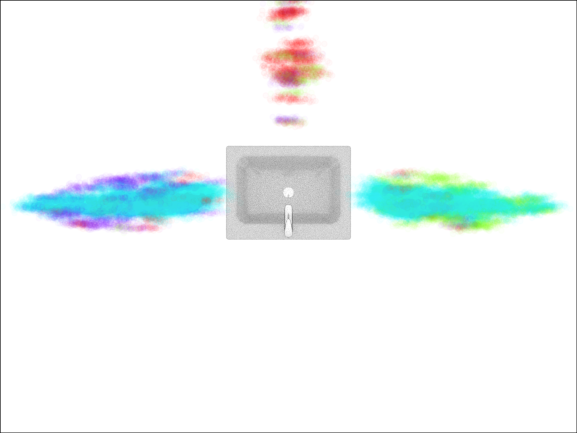}
		\caption{Sink-Toilet}
	\end{subfigure}
	
	\caption{Several results of learnt priors. }
	\label{fig:resultofdpc}
\end{figure*}

Patterns are priors suggesting how we arrange objects in real life layouts. Figure \ref{fig:pe_fitted} shows a pattern of a laptop \textit{w.r.t} an office chair. Since relative translations are incorporated, patterns can inherently avoid some unreasonable situations such as collisions. However, it is obvious that we cannot set a unified model for all patterns, since the patterns can have arbitrary shapes. 
To extract arbitrary-shaped patterns in discrete representations, we adopt the approach in
\cite{rodriguez2014clustering}, which clusters points according to $\rho$ (equation \ref{equ:localdensity}) and $\delta$ (equation \ref{equ:deltadis}), where the indicating function $I_{\{d \leq d_{c}\}}$ returns $1$ if $d \leq d_{c}$ and $0$ otherwise. 
\begin{align} 
\rho_{k} &= \sum_{k'} I_{\{d \leq d_{c}\}}(d_{k,k'}), d_{c} = d_{(\eta K^2)}, \label{equ:localdensity} \\ 
\delta_{k} &= \min_{j:\rho_{k} < \rho_{k'}}(d_{k,k'}). \label{equ:deltadis}
\end{align}
Given a set of edges $E^{i,j}_{in}$ from $v^{i}_{in}$ to $v^{j}_{in}$ in $G_{in}$ as shown in Figure \ref{fig:pe_origin}, we first calculate pairwise Euclidean distances between them using translations. For each edge $e^{i,j,k}_{in}$, a $\rho_{k}$ is counted as the number of other edges with distances less than $d_{c}$ to it. Taking $K$ points, $d_c$ is the $\eta K^2$- greatest value among all pairwise distances \hb{with $\eta = 0.015$} as suggested by \cite{rodriguez2014clustering}. $\delta_{k}$ represents the minimal distance from a set of $e^{i,j,k'}$ with higher $\rho_{k'}$ than $\rho_{k}$. 
As a result, despite arbitrary shapes, merely edges with high $\rho_{k}$ belong to a potential pattern, and each edge with high $\rho_{k}$ and high $\delta{k}$ indexes to a potential pattern, which is analogous to a cluster center in \cite{rodriguez2014clustering}. In contrast, noises tend to have high values of $\delta$ while their local density is distinctly low. As a result, we reduce noises and highlight patterns $E^{i,j}_{p}$, \hb{as illustrated in} figure \ref{fig:pe_denoised}. The rest of accurate patterns form a discrete templates $E_{p}^{i,j}$ are already fully usable to our framework. To incorporate our model in previous works, e.g., MCMC, our priors can be easily fitted to distributions such as using non-parametric kernel density estimation based on Gaussian kernels, as shown in Figures \ref{fig:pe_fitted} and \ref{fig:resultofdpc}.

We also \hb{perform similar prior extraction tasks} 
for individual objects with regard to their orientations and distances to the nearest walls where $d_{k,k'}$ becomes differences of scalars. In doing so, we keep the values $t_{w},\theta_{w}$ with both high values of $\rho$ and $\delta$ to index the pattern. Then we formulate the translation and rotation priors of walls into both multinomial distribution, and assign them to their corresponding vertices in $G_{in}$. 

\begin{figure}[!ht]
    \centering
    \includegraphics[width=\linewidth]{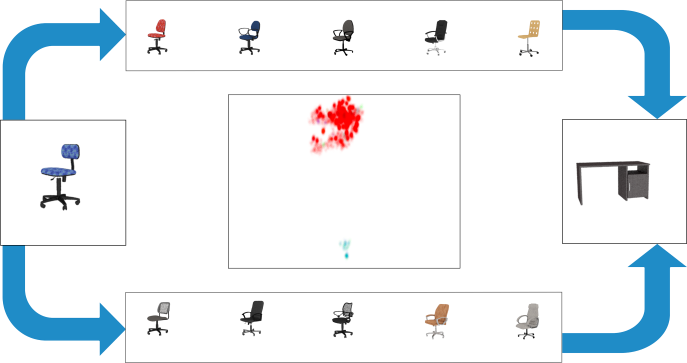}
    \caption{Assigning existing templates \hb{to new objects of similar geometry. Given a previously unseen office chair (Left), we achieve the layout strategy of it \textit{w.r.t} the desk (Right) by merging templates of objects geometrically similar to the chair (Middle). }}
    \label{fig:pe_general}
\end{figure}

Next, we further generalize our templates to make them reusable and extensible. We observed that objects with same semantics and similar geometries share layout strategies. As shown in Figure \ref{fig:pe_general}, given a new object \hb{without the corresponding priors extracted from our datasets}, we find its similar models by comparing 3D shapes of models using \cite{kleiman2015shed}, which uses $s_{shed}^{k}$ to measure the degree of similarity. We select the top-$K$ results $\{(o_{k}, s_{shed}^{k})|s_{shed}^{K} = \beta, k=1,2,3, \dots, K\}$ and take the union of the $K$ templates \hb{as the template for the new object}, where $\beta$ is chosen as $0.1$ according to our experiments.

Figure \ref{fig:resultofdpc} shows some results of Learnt priors. Similar to the visualization of dense optical flows \cite{farneback2003two}, we apply the system of hue, saturation and value (HSV) to represent orientations, where angles are normalized within $(0, 2\pi)$ as hue, probability densities are represented as saturation, and values are all set to $1$. Since height differences for most objects do not vary significantly, we plot the three channels $(p_x^{i,j,k},p_z^{i,j,k},p_\theta^{i,j,k})$ to make it more intuitive.

\section{Scene Synthesis} \label{sec:online}
In this section, we incorporate the learnt SSG and priors to synthesize room layouts. A synthesis process typically includes two steps: a heuristic arrangement, followed by an optimization. Given a set of input objects $\hat{O}$, we first decompose them into several groups according to the SSG, and arrange objects within each group, where relative transformations are immediately indexed by the templates. Finally, we apply a global optimization to satisfy layout strategies of objects in $\hat{O}$. 

\subsection{Heuristic Layouts with Formulated Groups} 
\begin{figure}[!ht]
	\centering
	\begin{subfigure}[b]{0.45\linewidth}
		\includegraphics[width=\linewidth]{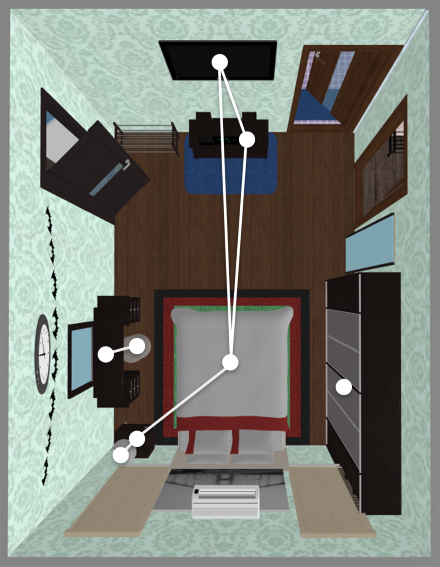}
		\caption{Marked relations. }
		\label{fig:srg_real}
	\end{subfigure}
	\hfill
	\begin{subfigure}[b]{0.45\linewidth}
		\includegraphics[width=\linewidth]{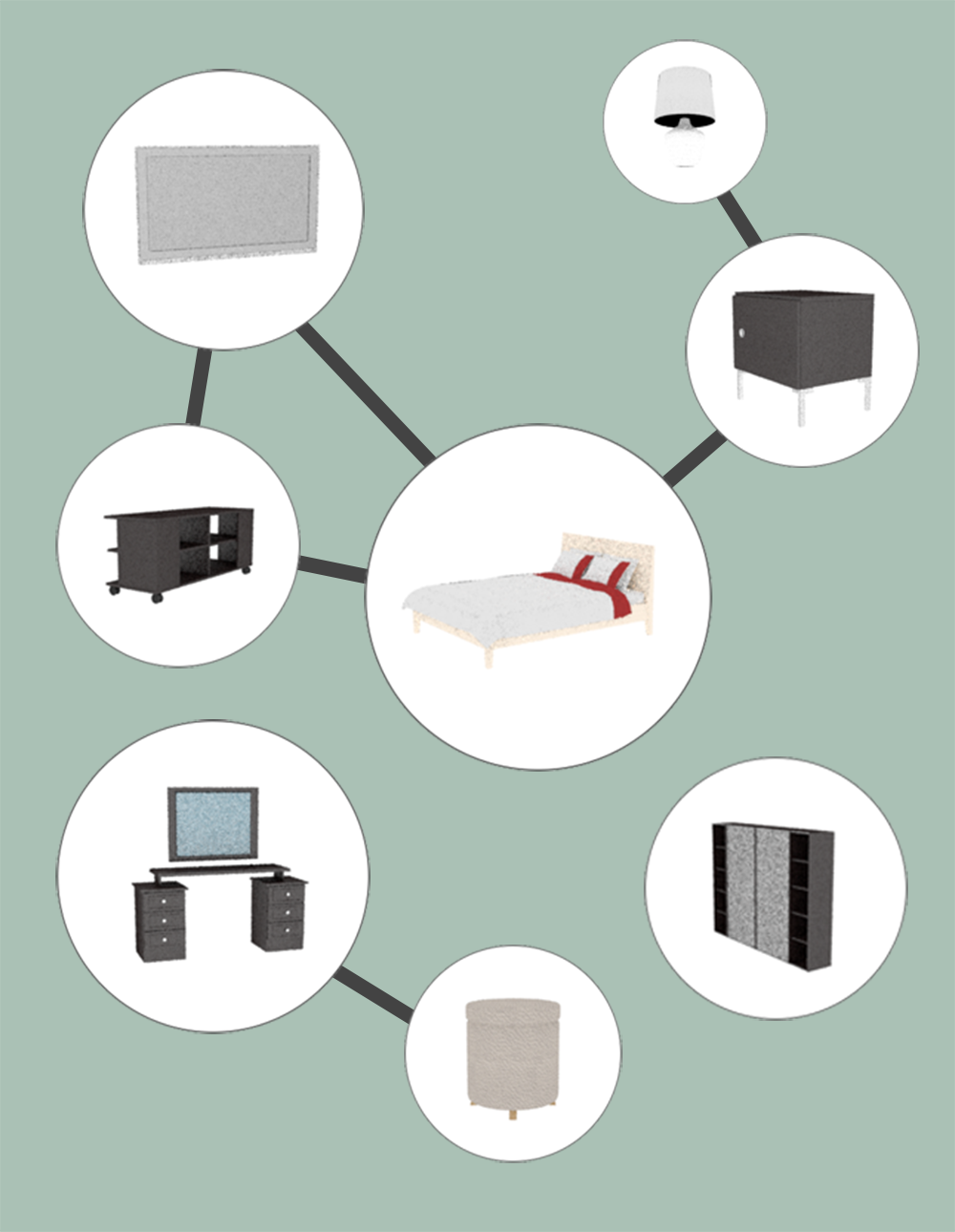}
		\caption{Disjoint graphs. }
		\label{fig:srg_abstract}
	\end{subfigure}
	\caption{Formulating functionally coherent groups of objects using tests for CSR. }
	\label{fig:group_relation}
\end{figure}

We first construct an unweighted graph described by an adjacency matrix $M_{adj}$, whose vertices correspond to input objects $\hat{O}$. Entries of $M_{adj}$ are determined by $G_{ss}$ in Section \ref{sec:csr}. More specifically, if $d^{u,v} \geq \epsilon$, then we set $M_{adj}^{u,v} = 1$, where $\epsilon$ typically equals to $1.628$ as suggested in \cite{assuncao1994testing}. After achieving $M_{adj}$, we iteratively construct disjoint groups $g \in Gr$ of objects by finding connected components of the graph represented by $M_{adj}$. Figure \ref{fig:group_relation} \hb{shows examples of resulting groups}. It is common to see a group containing only one object, such as wardrobe, cabinet or shelf, because their placement usually does not require considering other objects, 
\hb{Such single-object groups greatly ease the following optimization process.} 

Based on a given room shape, partitioned groups, and learnt templates, we then generate proposals for pending scenes, i.e., objects are immediately placed and oriented \textit{w.r.t} their groups and walls. For each group $g \in Gr$, layouts of $g$ are heuristically generated by sampling a posterior probability distribution $\Psi_{G|E_{p}}(g)$ expressed in Equation \ref{equ:ha}, given templates $E_p$ (Section \ref{sec:pe}). 
\begin{align} \label{equ:ha}
\Psi_{G|E_{p}}(g) &= \frac{\alpha(g) \cdot \Phi_{E_{p}|G=g}(E^\mu_p)}{\int \alpha(g) \cdot \Phi_{E_{p}|G=g}(E^\mu_p) dg}, \\ 
&= \frac
{\alpha(g) \cdot \sum_{\mu}\prod_{\tau \in g} \phi_{E_{p}^{\mu}|T=\tau}(E_{p}^{\mu,t},\tau^{\mu})}
{\int \alpha(g) \cdot \Phi_{E_{p}|G=g}(E^\mu_p) dg} \label{equ:ha_expand}, 
\end{align}
$\alpha(g)$ denotes the probability of each object $\tau \in g$ being the dominant object $\tau^{\mu}$ in $g$. \hb{Let $deg(\tau)$ denote} the degree of $\tau$ \textit{w.r.t} $M_{adj}$, which is the number of objects connected with it according to the test for CSR (Section \ref{sec:csr}), and $dmax = \max_{\tau \in g}deg(\tau)$. The likelihood $\phi_{E_{p}^{\mu}|T=\tau}(\cdot)$ is a multinomial distribution formed by the given template $E_{p}^{\mu,t}$ of $\tau$ \textit{w.r.t} $\tau^{\mu}$, while it is equal to a constant when $\tau = \tau^{\mu}$. 
\begin{equation} \label{equ:ha_prior}
\alpha(g)=
\begin{cases}
\frac{1}{|\{\tau|\tau \in g, deg(\tau) = dmax\}|}&, \text{ if } deg(\tau^{\mu}) = dmax \\
0&, \text{ otherwise}
\end{cases}
,
\end{equation}

When sampling $\Psi_{G|\Theta=\hat{O}}$, we first randomly decide $\tau^{\mu}$ of $g$. Equation \ref{equ:ha_expand} implies that $\{\phi_{E_{p}^{\mu}|T=\tau}(\cdot) | \tau \in g\}$ are independent to each other, so transformations of objects are sampled according to their own templates, respectively. In practice, if an object has a relatively low d-value to $\tau^{\mu}$, we further decompose the group and assign a new dominant object to it. In some cases, this heuristic strategy could sample a sufficiently plausible layout even without a further optimization. However, the heuristic strategy may still results in unreasonable conditions such as collision between groups, objects out of room boundaries, etc. Next we show how we adjust objects so that a plausible layout of objects is eventually presented. 

\subsection{Template Matching}
Equation \ref{equ:overall} mathematically formalizes template matching, where we are trying to minimize the summation of Hausdorff distances $d_{H}$ between all objects \textit{w.r.t} their templates. $X^{i}$ indexes the transformation of object $o^{i}$ and $E_{p}$ is a set of sampled transformations in Section \ref{sec:pe}. 
\begin{align} \label{equ:overall}
X^{*} &= \mathop{\arg\min}_{X} L(X,E_{p}) \\
&= \mathop{\arg\min}_{X} \sum_{i,j}M_{adj}^{i,j}d_{H}(X^{i}, E_{p}^{i,j}) + Col(X, r), 
\end{align}

$d_{H}$ is Hausdorff metric between an element to a set derived by the distance function $d_{h}$ under the space of translation and rotation. The reason for assembling Hausdorff distance is that it directly tackles samples instead of distributions. As illustrated, it is unlikely to mathematically express a unified distribution to model arbitrary layout patterns. In contrast, if we could extract samples of arbitrary shape, Hausdorff metric enables pipelines to skip model fitting and to optimize directly using refined samples. 
\begin{equation} \label{equ:hausdorff}
d_{H}(x, S) = \min_{v \in S} d_{h}(x, s), 
\end{equation}
\begin{align} 
d_{h}(x, s) &= \norm{x_p - v_p} + \exp(ori(x_{\theta}, v_{\theta})), \\
ori(\theta, \theta') &= \min (2\pi - \abs{\theta - \theta'}, \abs{\theta - \theta'}), 
\end{align}

Equation \ref{equ:collision} represents the artifacts among objects and between objects and walls, where $p(\chi, k)$ returns the $k$-th rotated point position of the bounding box or the shape of $\chi$. Ideally, if there is no collision and no object out of boundary, $Col(X, r)$ should equal to $0$. 
\begin{align}
Col(X, r) &= Col_{wall}(X, r) + Col_{obj}(X) \nonumber \\ 
&= \sum_{i,k} \prod_{r} tR(p(X_{i}, k), p(R, r), p(R, r + 1)) \nonumber \\
&+ \sum_{i,k,j} \prod_{l} tL(p(X_i, k), p(X_j, l), p(X_j,l + 1)) \label{equ:collision}
\end{align}
$Col_{wall}$ measures whether or not objects are out of walls, whilst $Col_{obj}$ calculates overlaps among objects. Truncated by $tR(\cdot)$ and $tL(\cdot)$, $\gamma(\cdot)$ represents the ``to-left" test of computational geometry \cite{de1997computational}, such as the utilization in \cite{graham1972efficient}. In addition to given objects, we place extra virtual objects as doors and windows with fixed transformations to avoid blocking them. 
\begin{align} \label{equ:toright}
tR(p_{1},p_{2},p_{3}) &= \max(-\gamma(p_{1},p_{2},p_{3}), 0) \\ 
tL(p_{1},p_{2},p_{3}) &= \max( \gamma(p_{1},p_{2},p_{3}), 0) \label{equ:toleft}
\end{align}

Since the underlying metrics are factorized as quadratic terms, we optimize equation \ref{equ:hausdorff} utilizing Position-Based Dynamics (PBD) \cite{bender2014survey}, which is also detailed in \cite{weiss2018fast}. Incorporating heuristic approaches, syntheses require $10$ iterations to converge on average after heuristic attempts. 

\section{Experiments}
\subsection{Tests for CSR}
\begin{figure}[!t]
	\centering
	\includegraphics[width=\linewidth]{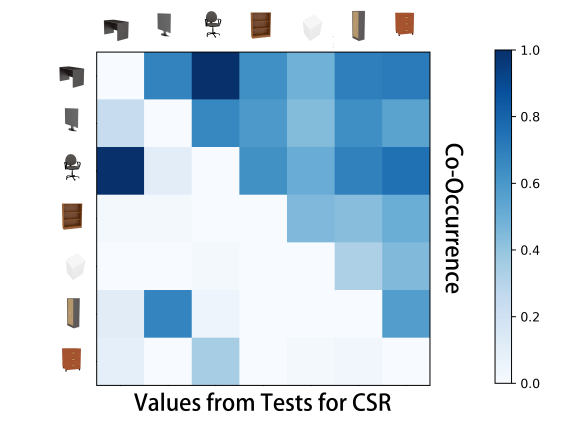}
	\caption{A comparison between values from tests for CSR and co-occurrence.}
	\label{fig:csrvscoo}
\end{figure}

Figure \ref{fig:csrvscoo} shows the comparison between using co-occurrence and using tests for CSR to measure the strengths of relations between objects. The results is normalized due to different scales. The upper triangular part depicts co-occurrence and the lower fills results from tests for CSR, which alleviates the unreasonableness caused by co-occurrence. The shelf is weakly related to the computer spatially, but co-occurrence suggests a strong relation.

It is obvious that placing the gray desk is independent of arranging the brown shelf, but they have a high frequency of co-existing in different rooms of various types, which potentially influences overall performances. Applying tests for CSR for them, the pair is decoupled spatially. The same is true of objects preferring independent layout with most of the others, such as the white dryer, the wardrobe and the brown stand. 

\subsection{Prior Sampling}
\begin{figure}[!ht]
	\centering
	\begin{subfigure}[b]{0.32\linewidth}
		\includegraphics[width=\linewidth]{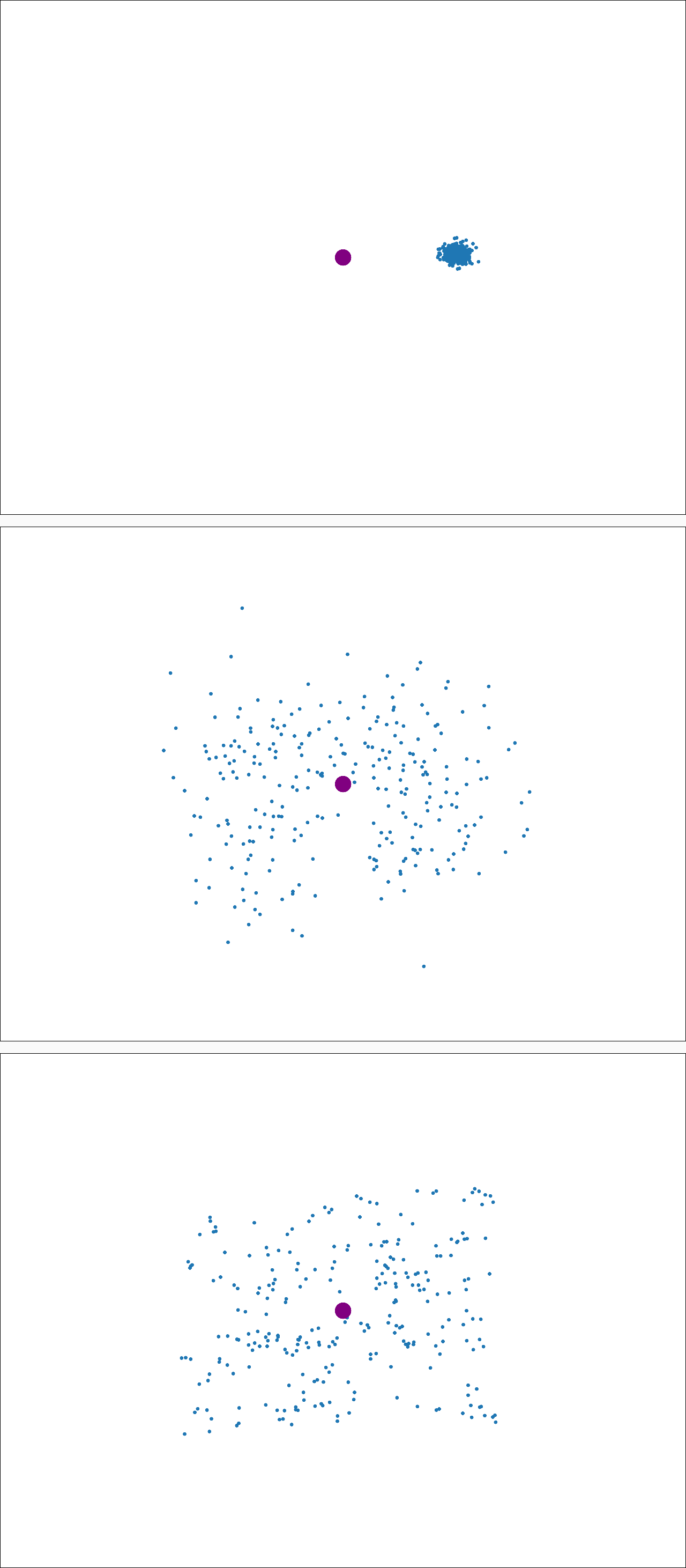}
		\caption{{\scriptsize DiningTable-Glass}}
		\label{fig:ps_diningtable_glass}
	\end{subfigure}
	\hfill
	\begin{subfigure}[b]{0.32\linewidth}
		\includegraphics[width=\linewidth]{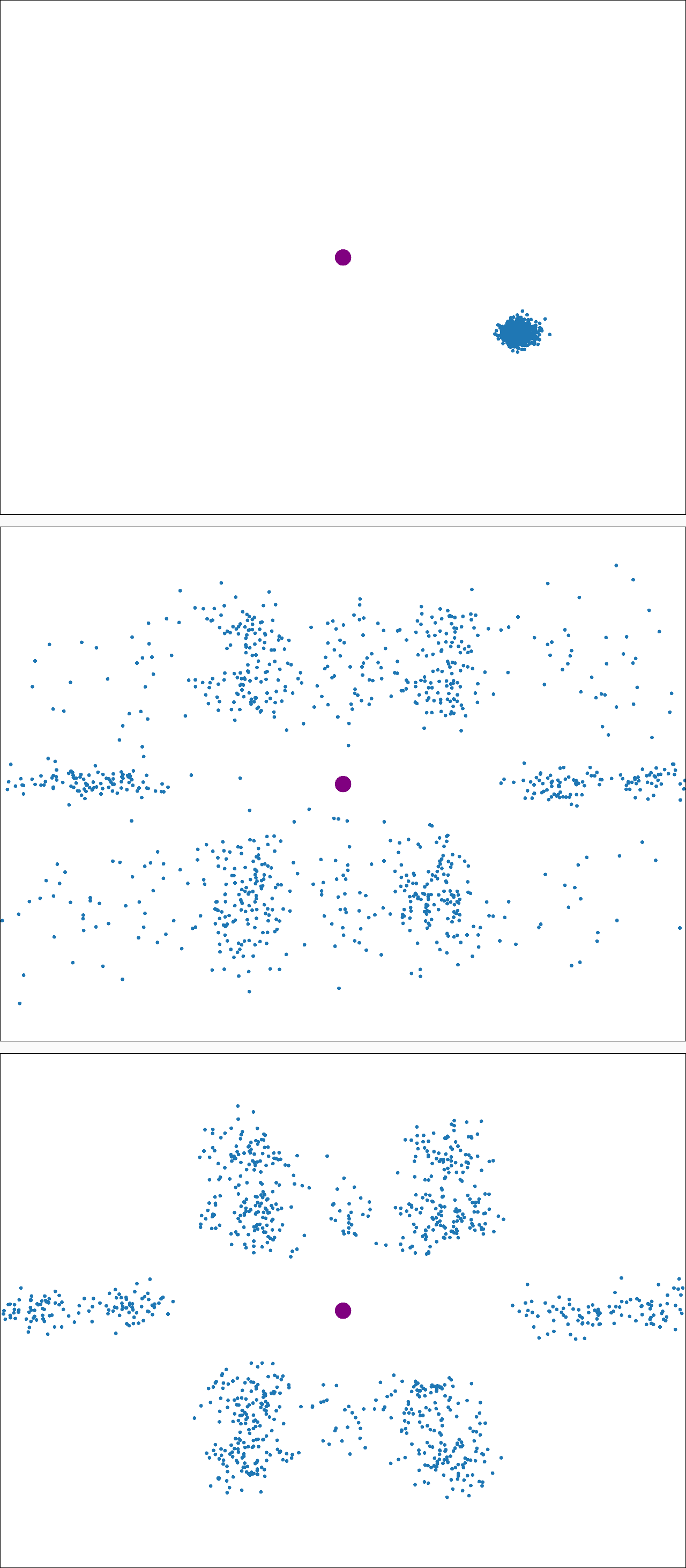}
		\caption{{\scriptsize LongTable-Chair}}
		\label{fig:ps_diningtable_chair}
	\end{subfigure}
	\hfill
    \begin{subfigure}[b]{0.32\linewidth}
		\includegraphics[width=\linewidth]{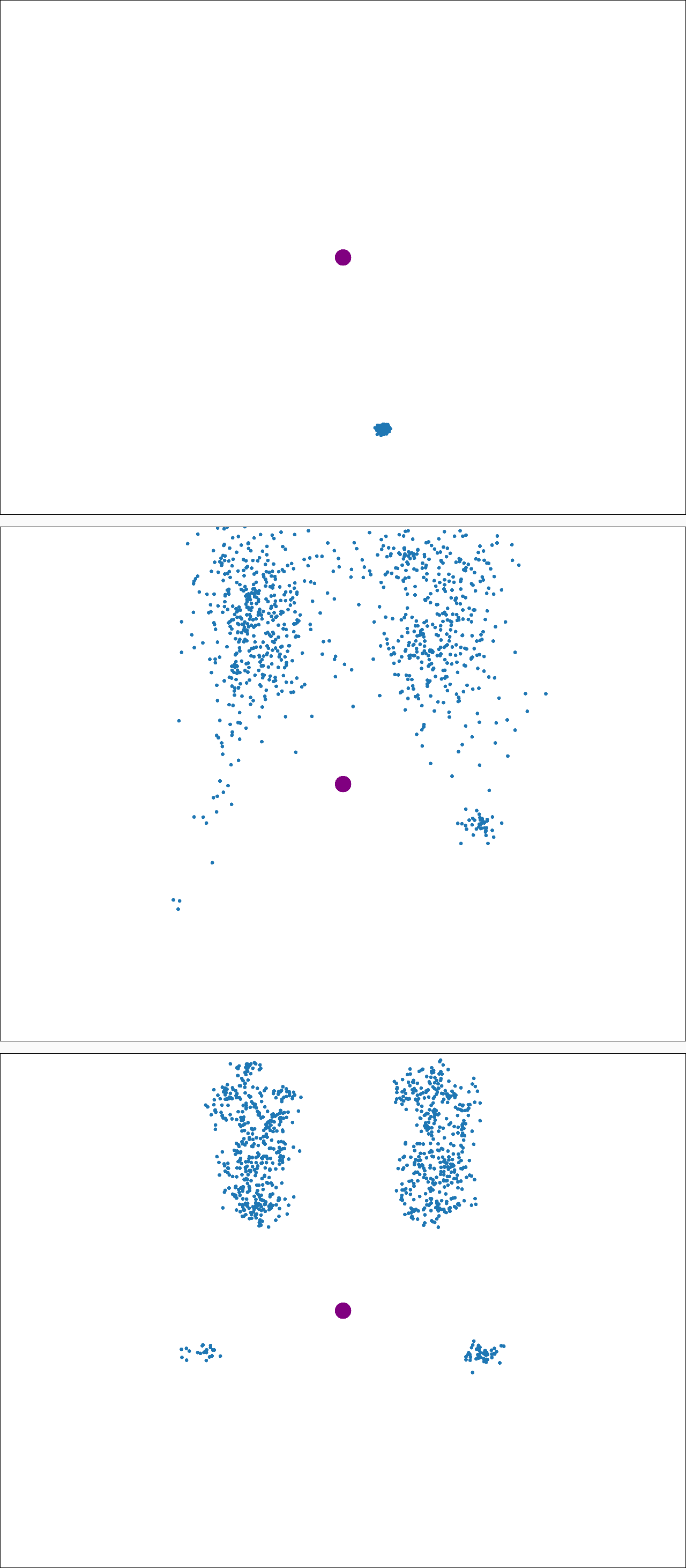}
		\caption{{\scriptsize Sofa-Loudspeaker}}
		\label{fig:ps_sofa_loudspeaker}
	\end{subfigure}
	
	\caption{Sampled transformations from extracted priors between objects, where the former is centred as red dots. Top: Yu et al. \cite{yu2011make}. Middle: Xu et al. \cite{xu2013sketch2scene}, Fisher el al. \cite{fisher2012example}. Bottom: Ours. }
	\label{fig:priorsampling}
\end{figure}

Figure \ref{fig:priorsampling} compares priors of our work with others. Since priors model layouts with high probability of being plausible between objects, we should get likely transformations by sampling them. According to experiments, both \cite{yu2011make} and GMM fail if noises exist in datasets, so inputs to all priors are data de-noised by us (section \ref{sec:pe}). Figure \ref{fig:priorsampling} shows the sampled transformations of priors, given de-noised data (section \ref{sec:pe}). 

Red dots denote the centred objects. The top row is priors used by \cite{yu2011make}, where they average the relative distances and orientations. We disturb their mean distance and orientations by a Gaussian kernel $\kappa \sim \mathcal{N}(0, 0.1)$. The middle row is GMM used by \cite{xu2013sketch2scene,fisher2012example}. Although we further manually set appropriate thresholds for each pattern in order to reduce potential noises as well as assembling \cite{akaike1998information} to assist explorations of number of peaks, the results are still confined to elliptical shapes such as figure \ref{fig:ps_diningtable_glass}, or introduced outliers such as \ref{fig:ps_diningtable_chair}. 
The bottom row shows our results. Ours are capable of detecting various layout patterns without introducing outliers. For example, figure \ref{fig:ps_sofa_loudspeaker} shows two patterns (four symmetry patterns) between a sofa and a loudspeaker. 

\subsection{Efficiency}
Our work achieves acceleration due to the usage of PBD \cite{bender2014survey} which is verified \cite{weiss2018fast} to be faster than using MCMC. However, our work is different from \cite{weiss2018fast} since ours is data-driven and does not require user input to constraint for each synthesis. In this section, we conduct several experiments to show the achieved efficiency, where examples are chosen from figure \ref{fig:results}. 

We do heuristic arrangement for both \cite{qi2018human} and \cite{yu2011make} to speed up their work. The time costs are shown in table \ref{tab:efficiency}, where values with ``greater-than signs" denote examples requiring more than $20000$ iterations. Experimentally, determining terminations for MCMC is hard and proposal moves are precarious. Because we judge whether or not a proposal is accepted \textbf{after} each iteration, resources are wasted. 

\begin{table}[!t]
	\caption{Time Consumption (sec). }
	\label{tab:efficiency}
	\centering
    \label{tab:table1}
    \setlength\tabcolsep{1.4pt}
    \begin{tabular}{ccccc} %
    \noalign{\hrule height 1.0pt}
    & \# Objects &Yu et al. \cite{yu2011make}&Qi et al. \cite{qi2018human}& Ours \\
    \hline
    Bedroom & 9 & $>$299.27s & 229.76s & 0.28s \\
    Living Room & 25 & $>$2135.30s & 1790.65s & 1.88s \\
    Bathroom & 8 & $>$313.54s & 216.20s & 0.29s \\
    Hybrid-1 & 13 & $>$714.60s & $>$481.35s & 0.64s \\
    Hybrid-2 & 35 & $>$2313.38s & $>$1667.03s & 2.31s \\
    Hybrid-3 & 28 & $>$1351.15s & 1122.63s & 1.24s \\
    \noalign{\hrule height 1.0pt}
    \end{tabular}
\end{table}

\subsection{User Study}
\begin{figure*}[!t]
	\centering
	\begin{subfigure}[b]{0.32\linewidth}
		\includegraphics[width=\linewidth, scale=0.5]{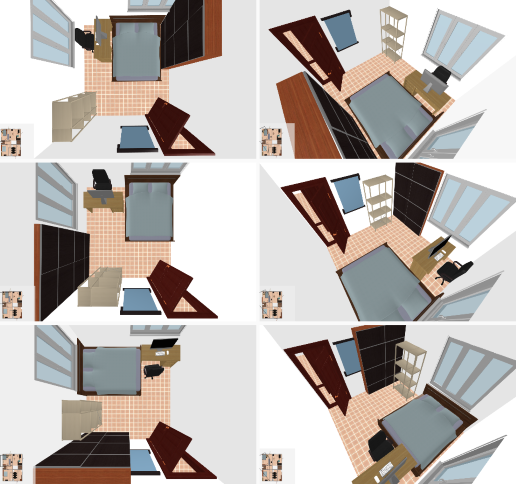}
		\caption{Bedroom}
	\end{subfigure}
	\hfill
	\begin{subfigure}[b]{0.32\linewidth}
		\includegraphics[width=\linewidth, scale=0.5]{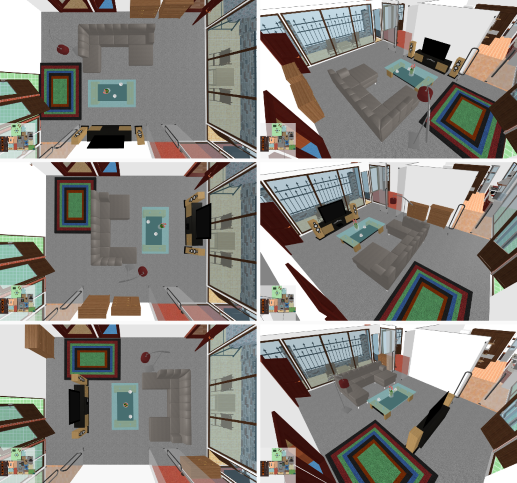}
		\caption{Living Room}
	\end{subfigure}
	\hfill
	\begin{subfigure}[b]{0.32\linewidth}
		\includegraphics[width=\linewidth, scale=0.5]{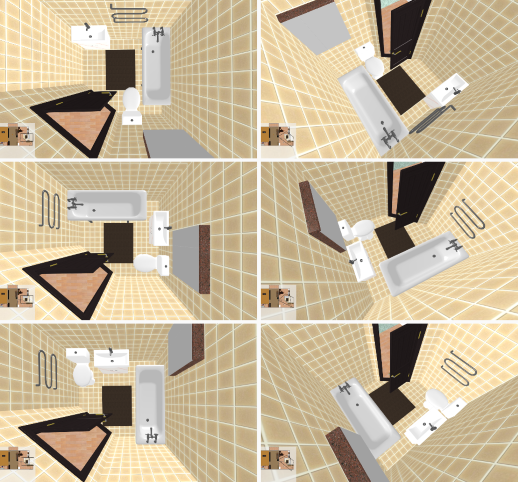}
		\caption{Bathroom}
	\end{subfigure}

	\begin{subfigure}[b]{0.32\linewidth}
		\includegraphics[width=\linewidth, scale=0.5]{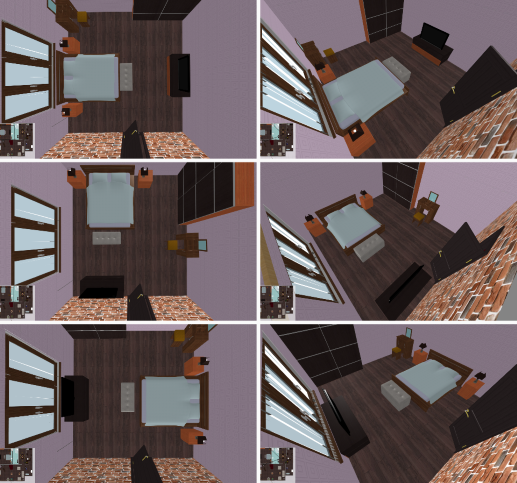}
		\caption{Hybrid Room - 1}
	\end{subfigure}
	\hfill
	\begin{subfigure}[b]{0.32\linewidth}
		\includegraphics[width=\linewidth, scale=0.5]{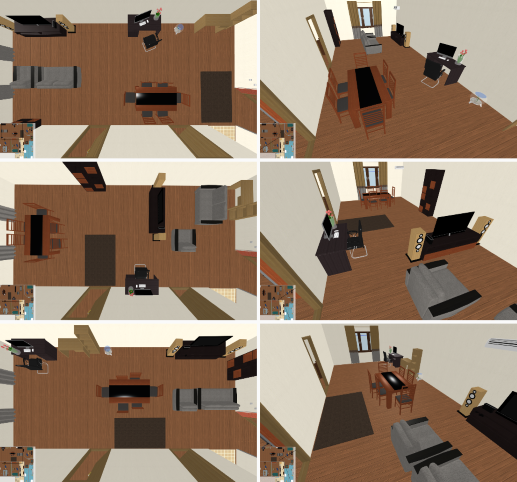}
		\caption{Hybrid Room - 2}
		\label{fig:result_kitchen_lr_sr}
	\end{subfigure}
	\hfill
	\begin{subfigure}[b]{0.32\linewidth}
		\includegraphics[width=\linewidth, scale=0.5]{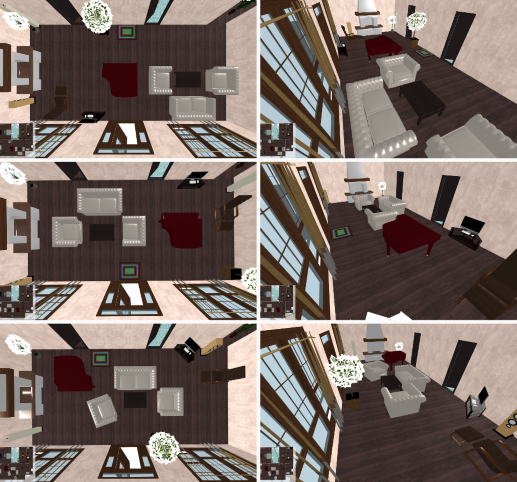}
		\caption{Hybrid Room - 3}
	\end{subfigure}
	
	\caption{Examples of various synthesized results. Each scene is generated with three alternative layouts in top views followed by side views. }
	\label{fig:results}
\end{figure*}

\begin{table*}[!ht]
    \centering
    \caption{User study: aesthetics. }
    \setlength\tabcolsep{3.6pt}
    \begin{tabular}{c|c|c|c|c|c|c|c|c|c}
    \noalign{\hrule height 1.0pt}
        Data & Bedroom & Living Room & Bathroom & Dinning Room & Balcony & Hall & Garage & \textbf{Hybrid Room} & Total \\
        \hline
        Ours & 2.944 & 3.292 & 2.989 & 3.344 & 3.344 & 3.061 & 3.256 & 3.317 & 3.194 \\
        \hline
        Ground Truth & 2.911 & 3.422 & 3.156 & 3.589 & 3.378 & 2.878 & 3.511 & 3.367 & 3.276 \\
    \noalign{\hrule height 1.0pt}
    \end{tabular}
    \label{tab:aesthetic}
\end{table*}
\begin{table*}[!ht]
    \centering
    \caption{User study: evaluations of tests for CSR and co-occurrence. }
    \setlength\tabcolsep{3.6pt}
    \begin{tabular}{c|c|c|c|c|c|c|c|c}
    \noalign{\hrule height 1.0pt}
        Metric & Bedroom & Living Room & Bathroom & Dinning Room & Balcony & Hall & Garage & Total \\
        \hline
        Tests for CSR & 93.31\% & 85.47\% & 96.67\% & 92.42\% & 86.36\% & 89.47\% & 76.17\% & 88.55\% \\
        \hline
        Co-Occurrence & 32.26\% & 43.81\% & 86.67\% & 45.76\% & 23.08\% & 38.46\% & 36.84\% & 43.53\% \\
    \noalign{\hrule height 1.0pt}
    \end{tabular}
    \label{tab:applicability}
\end{table*}
Results of our work is shown in figure \ref{fig:results}. Formulating functional groups using CSR enables us to generate hybrid rooms. Evaluations of 3D indoor scenes are subjective, so we conduct two user studies to evaluate our method. Firstly, aesthetic measures how visually pleasing the generated scenes are, i.e., asking subjects to grade generated layouts shuffled with ground truth. Subjects grade from level-1 (poor) to level-5 (perfect). As listed in table \ref{tab:aesthetic}, our generated results are comparable to the original layouts. Another user study is conducted to measure how tests for CSR and learnt templates satisfy intuitions of humans. We sort pairwise relations by tests for CSR and co-occurrence respectively. For each sorted list of pairs, we take templates of pairs at a fixed interval $int=120$ from the highest value. Then subjects judge whether or not the presented templates are consistent with real-life layout strategies. Tabulated in table \ref{tab:applicability}, results for co-occurrence contain considerable pairs spatially independent. In total, $63$ scenes and $500$ templates are generated. We invite $97$ subjects from societies and they were merely told to grade layouts and and judge patterns. 

\section{Conclusion}
In this paper, we present a framework for 3D indoor scene synthesis based on analysis of patterns. We experimentally verify the correctness, generalization and effectiveness of it. Our framework is capable of further expansion by easily incorporating object selections such as \cite{qi2018human,liang2017automatic}. Future work includes getting finer comparisons of 3D shapes for generalizing our templates such as 3DMatch \cite{zeng20163dmatch}. Recently, improvements for density peak clustering are also available \cite{tong2019density,liu2019constraint}. We hope the pipeline, learnt models and synthesized layouts can contribute to automatic room layouts as well as associated domains such as scene understanding \cite{satkin2012data}. 

{\small
\bibliographystyle{ieee_fullname}
\bibliography{3diss}
}

\end{document}